# Dynamics of two externally driven coupled quantum oscillators interacting with separate baths based on path integrals

Illarion Dorofeyev[*]

Institute for Physics of Microstructures, Russian Academy of Sciences,

603950, GSP-105 Nizhny Novgorod, Russia

## Abstract

The paper deals with the problem of dynamics of externally driven open quantum systems. Using the path integral methods we found an analytical expression for time-dependent density matrix of two externally driven coupled quantum oscillators interacting with different baths of oscillators. It is shown that at the zeroing of external forces the density matrix becomes identical to the previously obtained one for freely developing coupled oscillators. Mean values of observables are computed by using the Hermitian part of the matrix. All elements of the covariance matrix composed by coordinates and momenta of two driven coupled oscillators are calculated. The time-dependent mean values, dispersions and covariances of coordinates of coupled oscillators at given external forces are numerically studied. It is shown that the larger the coupling constant the larger is the disturbances of the second oscillator due to external action on the first oscillator. Coupled dynamics of forced oscillators at relatively large coupling constant is demonstrated at different thermodynamic conditions.

# I. Introduction

The physics of open systems covers various processes ranging from elementary-particle level to astrophysical scales. Also, such systems excite a great interest in technology and social sciences. An often-used model of open quantum systems consists of a quantum oscillator (set of oscillators) coupled to a heat reservoir (set of shared or separate reservoirs) of harmonic oscillators. This mechanical hamiltonian system allows investigating of dissipation, decoherence, correlation, continuous quantum measurement, quantum-to-classical transition and other important phenomena in nature. It is well-known that the model has been successful in describing the Brownian dynamics of selected particles coupled to a bath, see, for example, [1-19]. Study of the pure Hamiltonian composite system yields a reason of irreversibility in the dynamics of a selected quantum system interacting with surroundings after reduction with respect to reservoirs variables. It always deserves considerable attention because of transition from pure mechanical dynamics to thermodynamical laws due to inevitable effects of the environment on objects under study. Thus, the detailed description of relaxation processes of open systems to stationary or to quasi-stationary states, and to equilibrium or to quasi-equilibrium thermodynamic states is obviously very important.

For example, in the case of a harmonic oscillator with arbitrary damping and at arbitrary temperature an explicit expression for the time evolution of the density matrix when the system starts in a particular kind of pure state was derived and investigated in a seminal work [20] based on the path integral technique. It was shown that the spatial dispersion in the infinite time limit agrees with the fluctuation-disspation theorem (FDT). To study the approach to equilibrium or to some transient stationary state of the system, a problem for coupled oscillators interacting with heat baths characterized by its own temperatures was considered in [21-26]. It was concluded that an arbitrary initial state of a harmonic oscillator state decayes towards a stationary state. A study of a harmonic chain to whose ends independent heat baths are attached which kept at different temperatures was performed in [27]. It was found the chain approaches a stationary state regime. An analyze of the nonequilibrium steady states of a one-dimensional harmonic chain of atoms with alternating masses connected to heat reservoirs at unequal temperatures and a heat transfer across an arbitrary classical harmonic network connected to two heat baths at different temperatures were done in [28, 29]. The evolution of quantum states of networks of quantum oscillators coupled with arbitrary external environments was analyzed in [30]. The emergence of thermodynamical laws in the long time regime and some constraints on the low frequency behavior of the environmental spectral densities were demonstrated.

Much attention has been focused on the bipartite continuous variable systems composed of two interacting oscillators.

Based on the non-Markovian master equations, the entanglement evolution of two harmonic oscillators under the influence of non-Markovian thermal environments was studied in [31]. It was demonstrated that the dynamics of the quantum entanglement is depended on the initial states, the coupling between oscillators and the coupling to shared baths or to separated baths. A study of the time-dependent entanglement and quantum discord between two oscillators coupled to a common environment was provided in papers [32-34]. Different stages of evolution including sudden disappearance and appearance of the entanglement were described providing a characterization of this process for different reservoirs including Ohmic, sub-Ohmic, and super-Ohmic models. Two coupled oscillators in a common environment at arbitrary temperature and the quantum decoherence of their states were investigated in [35]. It was shown that the problem can be mapped into that of a single harmonic oscillator in a general environment plus a free harmonic oscillator. Besides, simplest cases of the entanglement dynamics were considered analytically and an analytical criterion for the finite-time disentanglement was derived at the Markovian approximation. The time-dependent entanglement between two coupled different oscillators within a common bath and within two separate baths was studied in [36] based on a master equation. It was found that in the case of separate baths at not very low temperatures the initial two-mode squeezed state becomes separable accompanying with a series of features. For instance, if the two oscillators share a common bath, the observation of asymptotic entanglement at relevant temperatures becomes possible. The evolution of quantum correlations of entangled two-mode states in a single-reservoir and in a two-reservoir model was studied in [37]. It was shown that in the two-reservoir model the initial entanglement is completely lost, and both modes are finally uncorrelated, but in a common reservoir both modes interact indirectly via the same bath. In [38] a system of two coupled oscillators within separate reservoirs was investigated. It was shown that if the baths are at different temperatures, then the interaction between the particles must be strong enough in order to reach a steady state entanglement. No thermal entanglement between two coupled oscillators is found in the high-temperature regime and weak coupling limits in [39]. The existence of a nonequilibrium state for two coupled, parametrically driven, dissipative harmonic oscillators which has stationary entanglement at high temperatures was reported in [40]. Based on exact results for the non-Markovian dynamics of two parametrically coupled oscillators in contact to independent thermal baths, the out-of-equilibrium quantum limit derived in [40] is generalized to the non-Markovian regime in [41]. It is shown that non-Markovian dynamics allows for the survival of stationary entanglement at higher temperatures. A stationary regime of two coupled oscillators connecting with independent reservoirs of harmonic oscillators was studied in [42], and analytical formulas for the mean energy of interaction of the selected oscillators and their mean energies in this case were derived. Time-dependent behavior of variances and covariances of two coupled oscillators within separate baths in the weak-coupling limit was investigated in [43]. It was demonstrated that these characteristics of two weakly coupled oscillators in the infinite time limit agrees with the FDT despite of initial variances. The case of arbitrary coupling of identical oscillators was considered in [44], and it was shown that the larger a difference in temperatures of thermal baths, the larger is a difference of the stationary values of variances of coupled identical oscillators as compared to values given by the FDT. The general case of two arbitrary coupled oscillators of arbitrary properties interacting with separate reservoirs is studied in [45]. As well as in previous cases the temporal dynamics of spatial variances and covariances of oscillators from any given time up to quasi-equilibrium steady states is studied based on path integration. It is shown for arbitrary oscillators that the spatial variances and covariances achieve stationary values in the long-time limit. It is demonstrated that the larger the difference in masses and eigenfrequencies of coupled oscillators, the smaller are the deviations of stationary characteristics from those given by the FDT at fixed coupling strength and fixed difference in temperatures between thermal baths.

The main goal of this paper is to derive an analytical expression for a temporary dependent density matrix of two selected oscillators subjected by two independent external forces at any arbitrary times. The reduced density matrices allow calculating whole set of elements of a covariance matrix. Temporal behavior of some mean values of observables is provided.

The paper is organized as follows. In Sec.II we describe a theoretical basis using a path integration method to calculate the reduced density matrix of two interacting quantum oscillators in different reservoirs of harmonic oscillators. Temporal dynamics and stationary states of mean values of coordinates are given in Sec.III. Our conclusions are given in Sec.IV.

## II. Problem statement and solution

We consider two bilinear coupled oscillators. In its turn, each of these oscillators is bilinear coupled with separate reservoirs of oscillators and subjected by external forces. Corresponding time-dependent Hamiltonian is written as follows

$$H(t) = p_1^2/2M_1 + M_1\omega_{01}^2 x_1^2/2 + p_2^2/2M_2 + M_2\omega_{02}^2 x_2^2/2 - \lambda x_1 x_2 - x_1 f_1(t) - x_2 f_2(t)$$
$$+ \sum_{j=1}^{N_1}\left[p_j^2/2m_j + m_j\omega_j^2(q_j - x_1)^2/2\right] + \sum_{k=1}^{N_2}\left[p_k^2/2m_k + m_k\omega_k^2(q_k - x_2)^2/2\right], \quad (1)$$

where $x_{1,2}, p_{1,2}, M_{1,2}, \omega_{01,02}$ are the coordinates, momenta, masses and eigenfrequencies of the selected oscillators, $\lambda$ is the coupling constant, $q_j, p_j, \omega_j, m_j$ and $q_k, p_k, \omega_k, m_k$ are the coordinates, momenta, eigenfrequencies and masses of bath's oscillators. Further we use the vectors $\vec{R}_1 = \{q_j\} = \{q_1,...,q_{N_1}\}$ and $\vec{R}_2 = \{q_k\} = \{q_1,...,q_{N_2}\}$ for brevity.

We suppose that, in the time $t = 0$ all interactions among oscillators are switched on and maintained during arbitrary time interval up to infinity. Then, the arbitrary external forces begin acting at arbitrary time moments $t_{01} > 0$, $t_{02} > 0$. The problem is to find a time-dependent density matrix of two driven coupled oscillators in any moment of time $t \geq 0$ followed by calculating all elements of a covariance matrix in this case. Figure 1 illustrates the above described scenario.

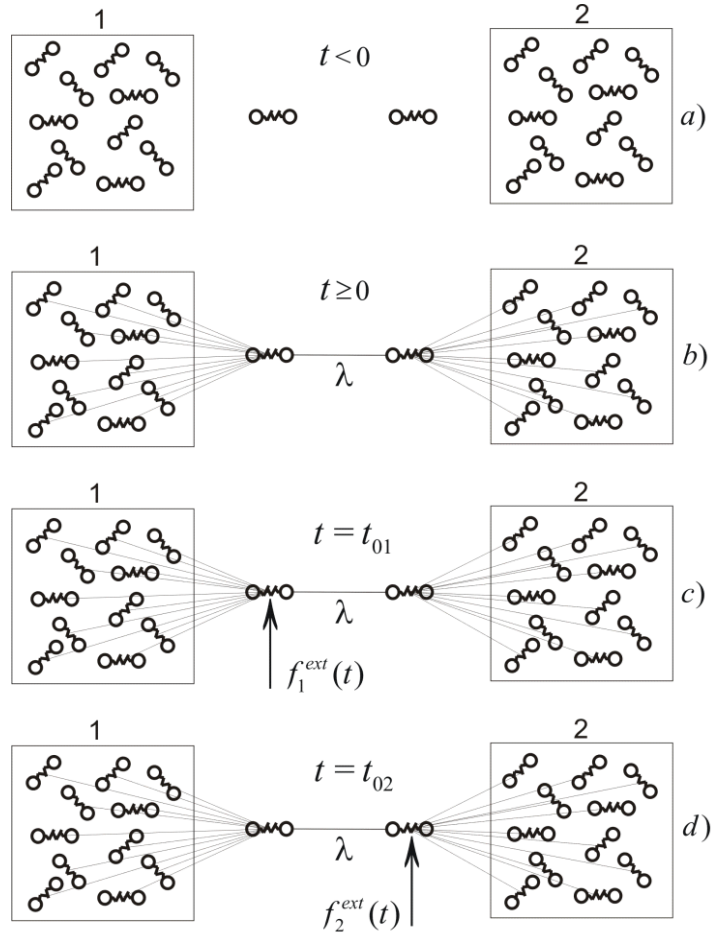

Fig.1

It is well known [46-52] that the evolution of the total Hamiltonian system "selected two interacting oscillators plus two reservoirs plus external forces" is described by the equation for the density matrix $W(t)$ of the total system

$$W(t) \equiv W_t = \hat{T}\exp\left[-(i/\hbar)\int_0^t H(s)ds\right]W(0)\hat{T}\exp\left[(i/\hbar)\int_0^t H(s)ds\right], \qquad (2)$$

where $W(0) \equiv W_0$ is the initially prepared density matrix of the total system, and we take into account the time dependence of the total Hamiltonian in Eq.(1) due to external forces, $\hat{T}$ is the time-ordering operator.

Further, we use the designations $\vec{x} = \{x_1, x_2\}$ and $\vec{R} = \{\vec{R}_1, \vec{R}_2\}$ for shorten notation. Using these designations we write the completeness property of the position eigenfunctions in the coordinate representation as usual

$$\int dx_1 dx_2 d\vec{R}_1 d\vec{R}_2 \left|x_1 x_2 \vec{R}_1 \vec{R}_2\right\rangle\left\langle x_1 x_2 \vec{R}_1 \vec{R}_2\right| \equiv \int d\vec{x}d\vec{R}\left|\vec{x}\vec{R}\right\rangle\left\langle\vec{x}\vec{R}\right| = 1, \qquad (3)$$

where the limits of the multiple integration extended from minus to plus infinity. Using Eq.(3), the Eq. (2) can be written in the matrix form as follows

$$\left\langle\vec{x}\vec{R}|W_t|\vec{y}\vec{Q}\right\rangle = \int d\vec{x}'d\vec{R}'d\vec{y}'d\vec{Q}'\left\langle\vec{x}\vec{R}|\hat{U}(t,0)|\vec{x}'\vec{R}'\right\rangle\left\langle\vec{x}'\vec{R}'|W_0|\vec{y}'\vec{Q}'\right\rangle\left\langle\vec{y}'\vec{Q}'|\hat{U}^\dagger(t,0)|\vec{y}\vec{Q}\right\rangle, \qquad (4)$$

where the unitary operator of evolution

$$\hat{U}(t,0) = \hat{T}\exp\left[-(i/\hbar)\int_0^t H(s)ds\right], \qquad (5)$$

The transition amplitudes in Eq.(4) are expressed via the path integrals [46-52] in different ways, for example, in the coordinate form

$$\left\langle\vec{x}\vec{R}|\hat{U}(t,0)|\vec{x}'\vec{R}'\right\rangle \equiv K(\vec{x},\vec{R},t;\vec{x}',\vec{R}',0) = \\ \int \mathsf{D}\underline{\bar{x}}_1 \mathsf{D}\underline{\bar{x}}_2 \mathsf{D}\underline{\vec{R}}_1 \mathsf{D}\underline{\vec{R}}_2 \exp\{(i/\hbar)S[\bar{x}_1(\tau),\bar{x}_2(\tau),\vec{R}_1(\tau),\vec{R}_2(\tau)]\}, \qquad (6)$$

where the integration along all paths is carried out from $\bar{x}_1(0) = x_1'$ to $\bar{x}_1(t) = x_1$, from $\bar{x}_2(0) = x_2'$ to $\bar{x}_2(t) = x_2$, and from $\vec{R}_1(0) = \vec{R}_1'$ to $\vec{R}_1(t) = \vec{R}_1$, from $\vec{R}_2(0) = \vec{R}_2'$ to $\vec{R}_2(t) = \vec{R}_2$. For convenience, we designate the integration variables by straight lines beneath and over the letters for the bath's and oscillator's coordinates, correspondingly.

The backward amplitude in Eq. (4) is

$$\left\langle\vec{y}'\vec{Q}'|\hat{U}^\dagger(t,0)|\vec{y}\vec{Q}\right\rangle \equiv K^*(\vec{y},\vec{Q},t;\vec{y}',\vec{Q}',0) = \\ \int \mathsf{D}\bar{y}_1 \mathsf{D}\bar{y}_2 \mathsf{D}\vec{Q}_1 \mathsf{D}\vec{Q}_2 \exp\{(-i/\hbar)S[\bar{y}_1(\tau),\bar{y}_2(\tau),\vec{Q}_1(\tau),\vec{Q}_2(\tau)]\}, \qquad (7)$$

where the integration along all paths is carried out from $\bar{y}_1(0) = y'_1$ to $\bar{y}_1(t) = y_1$, from $\bar{y}_2(0) = y'_2$ to $\bar{y}_2(t) = y_2$, and from $\vec{\bar{Q}}_1(0) = \vec{Q}'_1$ to $\vec{\bar{Q}}_1(t) = \vec{Q}_1$, from $\vec{\bar{Q}}_2(0) = \vec{Q}'_2$ to $\vec{\bar{Q}}_2(t) = \vec{Q}_2$.

The path integrals are expressed via the multiple Reihmann integrals in the coordinate space as usual [46-52]. The action $S[x_1, x_2, \vec{R}_1, \vec{R}_2]$ in Eq.(6) is expressed via the Lagrangian corresponding to the Hamiltonian in Eq.(1). This action in Eqs.(6) and (7) is structured in according with Eq.(1)

$$S(t) = S_{A1}(t) + S_{B1}(t) + S_{I1}(t) + S_{A2}(t) + S_{B2}(t) + S_{I2}(t) + S_{12}(t) + S_{ext}^{(1)}(t) + S_{ext}^{(2)}(t). \tag{8}$$

where

$$S_{Ai} = \int_0^t d\tau \left[ M_i \dot{x}_i^2(\tau)/2 - M_i \omega_{0i}^2 x_i^2(\tau)/2 \right], \quad (i=1,2), \quad S_{12} = \int_0^t dt' [\lambda x_1(\tau) x_2(\tau)],$$

$$S_{B1} = \int_0^t d\tau \sum_{j=1}^{N_1} (m_j/2)[\dot{q}_j^2(\tau) - \omega_j^2 q_j^2(\tau)], \quad S_{B2} = \int_0^t d\tau \sum_{k=1}^{N_2} (m_k/2)[\dot{q}_k^2(\tau) - \omega_k^2 q_k^2(\tau)],$$

$$S_{I1} = \int_0^t d\tau \sum_{j=1}^{N_1} [c_j q_j(\tau) x_1(\tau) - (c_j^2/2m_j \omega_j^2) x_1^2(\tau)], \tag{9}$$

$$S_{I2} = \int_0^t d\tau \sum_{k=1}^{N_2} [c_k q_k(\tau) x_2(\tau) - (c_k^2/2m_k \omega_k^2) x_2^2(\tau)],$$

$$S_{ext}^{(i)} = \int_0^t d\tau [x_i(\tau) f_i(\tau)], \quad (i=1,2)$$

Putting in this equation $c_j = m_j \omega_j^2$, $c_k = m_k \omega_k^2$ we obtain a system exactly corresponding to Hamiltonian in Eq. (1).

The reduced density matrix comprising only variables of selected oscillators is obtained by tracing the whole density matrix in Eq.(4) over the variables of the two baths

$$\rho(x_1, x_2, y_1, y_2, t) = \int d\vec{R} \lim_{\vec{Q} \to \vec{R}} \langle \vec{x}\vec{R} | W_t | \vec{y}\vec{Q} \rangle$$
$$= \int d\vec{x}' d\vec{R}' d\vec{y}' d\vec{Q}' \int d\vec{R} \, K(\vec{x}, \vec{R}, t; \vec{x}', \vec{R}', 0) \langle \vec{x}'\vec{R}' | W_0 | \vec{y}'\vec{Q}' \rangle K^*(\vec{y}, \vec{R}, t; \vec{y}', \vec{Q}', 0) \tag{10}$$

where transition amplitudes $K$ and $K^*$ from Eq.(6), (7), and $\langle \vec{x}', \vec{R}' | W_0 | \vec{y}', \vec{Q}' \rangle \equiv W(\vec{x}', \vec{y}'; \vec{R}', \vec{Q}'_2, 0)$ is the density matrix of the global system at the initial time $t = 0$.

### III. Results and discussion
#### A. Propagator for driven coupled oscillators

We choose the initial whole density matrix as follows

$$W(x'_1, x'_2, y'_1, y'_2; \vec{R}'_1, \vec{R}'_2, \vec{Q}'_1, \vec{Q}'_2, 0) = \rho_A^{(1)}(x'_1, y'_1, 0) \rho_A^{(2)}(x'_2, y'_2, 0) \rho_B^{(1)}(\vec{R}'_1, \vec{Q}'_1, 0) \rho_B^{(2)}(\vec{R}'_2, \vec{Q}'_2, 0), \tag{11}$$

where $\rho_A^{(1)}(x_1', y_1', 0)$, $\rho_A^{(2)}(x_2', y_2', 0)$ are initially prepared density matrices of the two selected oscillators, $\rho_B^{(1)}(\vec{R}_1', \vec{Q}_1', 0)$, $\rho_B^{(2)}(\vec{R}_2', \vec{Q}_2', 0)$ are initial density matrices of the separate reservoirs. In this case the reduced density matrix looks like

$$\rho(x_1, x_2, y_1, y_2, t) = \int dx_1' dx_2' dy_1' dy_2' \, J(x_1, x_2, y_1, y_2, t; x_1', x_2', y_1', y_2', 0) \rho_A^{(1)}(x_1', y_1', 0) \rho_A^{(2)}(x_2', y_2', 0) \,, \quad (12)$$

where the propagator

$$\begin{aligned}J(x_1, x_2, y_1, y_2, t; x_1', x_2', y_1', y_2', 0) = \int \mathsf{D}\bar{x}_1 \mathsf{D}\bar{x}_2 \mathsf{D}\bar{y}_1 \mathsf{D}\bar{y}_2 \, \exp\{(i/\hbar)\bigl(S_{A1}[\bar{x}_1] - S_{A1}[\bar{y}_1] \\ + S_{A2}[\bar{x}_2] - S_{A2}[\bar{y}_2] + S_{12}[\bar{x}_1, \bar{x}_2] - S_{12}[\bar{y}_1, \bar{y}_2] + S_{ext}^{(1)}[\bar{x}_1] - S_{ext}^{(1)}[\bar{y}_1] \\ + S_{A2}[\bar{x}_2] - S_{A2}[\bar{y}_2] + S_{12}[\bar{x}_1, \bar{x}_2] - S_{12}[\bar{y}_1, \bar{y}_2] + S_{ext}^{(2)}[\bar{x}_2] - S_{ext}^{(2)}[\bar{y}_2]\bigr)\} \mathsf{F}_{FV}[\bar{x}_1, \bar{x}_2, \bar{y}_1, \bar{y}_2]\end{aligned} \quad , \quad (13)$$

where all actions are defined by Eq.(9) and the integration along all paths is carried out from $\bar{x}_1(0) = x_1'$ to $\bar{x}_1(t) = x_1$, from $\bar{x}_2(0) = x_2'$ to $\bar{x}_2(t) = x_2$, and from $\bar{y}_1(0) = y_1'$ to $\bar{y}_1(t) = y_1$, from $\bar{y}_2(0) = y_2'$ to $\bar{y}_2(t) = y_2$, correspondingly.

The Feynman-Vernon influence functional $\mathsf{F}_{FV}$ in Eq.(13) is the same as in [43]. In our case we have

$$\mathsf{F}_{FV}[\bar{x}_1, \bar{x}_2, \bar{y}_1, \bar{y}_2] = \exp\left\{-\left(\Phi_{EV}^{(1)}[\bar{x}_1, \bar{y}_1] + \Phi_{EV}^{(2)}[\bar{x}_2, \bar{y}_2]\right)\right\}, \quad (14)$$

where

$$\begin{aligned}\Phi_{EV}^{(1)}[\bar{x}_1, \bar{y}_1] + \Phi_{EV}^{(2)}[\bar{x}_2, \bar{y}_2] = \hbar^{-1}\int_0^t dt' \int_0^{t'} dt'' \{\bar{x}_1(t') - \bar{y}_1(t')\}\{\alpha_1(t'-t'')\bar{x}_1(t'') - \alpha_1^*(t'-t'')\bar{y}_1(t'')\} \\ + i(\mu_1/2\hbar)\int_0^t dt'[\bar{x}_1^2(t') - \bar{y}_1^2(t')] + \\ \hbar^{-1}\int_0^t dt' \int_0^{t'} dt'' \{\bar{x}_2(t') - \bar{y}_2(t')\}\{\alpha_2(t'-t'')\bar{x}_2(t'') - \alpha_2^*(t'-t'')\bar{y}_2(t'')\} \\ + i(\mu_2/2\hbar)\int_0^t dt'[\bar{x}_2^2(t') - \bar{y}_2^2(t')]\end{aligned} \quad , \quad (15)$$

where

$$\mu_1 = \sum_{j=1}^N \frac{c_j^2}{m_j \omega_j^2} = \frac{2}{\pi}\int_0^\infty d\omega \, \frac{J_1(\omega)}{\omega}, \quad \mu_2 = \sum_{k=1}^N \frac{c_k^2}{m_k \omega_k^2} = \frac{2}{\pi}\int_0^\infty d\omega \, \frac{J_2(\omega)}{\omega}, \quad (16)$$

where $J_{1,2}(\omega)$ are the spectral densities of noise of two baths.

The functions $\alpha_{1,2}$ can be represented in the following convenient form

$$\alpha_{1,2}(t) \equiv \alpha_{1,2}'(t) + i\alpha_{1,2}''(t) = \frac{1}{\pi}\int_0^\infty d\omega \, J_{1,2}(\omega)[\coth(\omega\hbar\beta_{1,2}/2)\cos(\omega t) - i\sin(\omega t)], \quad (17)$$

see, for example [15].

We use the Ohmic case when $J_{1,2}(\omega) = \eta_{1,2}\omega$, which gives $\alpha_{1,2}''(t) \approx \eta_{1,2}\partial\delta(t)/\partial t$, where $\delta(t)$ is the delta-function, $\mu_{1,2} = 2\eta_{1,2}\nu_{1,2}^{max}/\pi$, where $\nu_{1,2}^{max}$ is the maximal frequency of excitations within baths. Taking into account Eq. (17) the Eq. (13) for propagator can be rewritten as follows

$$
\begin{aligned}
J(x_1, x_2, y_1, y_2, t; x_1', x_2', y_1', y_2', 0) = \int \mathsf{D}\bar{x}_1 \mathsf{D}\bar{y}_1 \exp\frac{i}{\hbar}\{S_{A1}[\bar{x}_1] - S_{A1}[\bar{y}_1] + S_{ext}^{(1)}[\bar{x}_1] - S_{ext}^{(1)}[\bar{y}_1] \\
-\int_0^t dt' \int_0^{t'} dt'' \{\bar{x}_1(t') - \bar{y}_1(t')\}\alpha_1''(t'-t'')\{\bar{x}_1(t'') + \bar{y}_1(t'')\}\} \\
\times \exp-\frac{1}{\hbar}\int_0^t dt'\int_0^{t'} dt''\{\bar{x}_1(t')-\bar{y}_1(t')\}\alpha_1'(t'-t'')\{\bar{x}_1(t'')-\bar{y}_1(t'')\}\exp-i\frac{\mu_1}{2\hbar}\int_0^t dt'[\bar{x}_1^2(t')-\bar{y}_1^2(t')] \\
\times \int \mathsf{D}\bar{x}_2 \mathsf{D}\bar{y}_2 \exp\frac{i}{\hbar}\{S_{A2}[\bar{x}_2] - S_{A2}[\bar{y}_2] + S_{ext}^{(2)}[\bar{x}_1] - S_{ext}^{(2)}[\bar{y}_1] + S_{12}[\bar{x}_1, \bar{x}_2] - S_{12}[\bar{y}_1, \bar{y}_2] - \\
-\int_0^t dt' \int_0^{t'} dt'' \{\bar{x}_2(t')-\bar{y}_2(t')\}\alpha_2''(t'-t'')\{\bar{x}_2(t'')+\bar{y}_2(t'')\} \\
\times \exp-\frac{1}{\hbar}\int_0^t dt'\int_0^{t'} dt''\{\bar{x}_2(t')-\bar{y}_2(t')\}\alpha_2'(t'-t'')\{\bar{x}_2(t'')-\bar{y}_2(t'')\}\exp-i\frac{\mu_2}{2\hbar}\int_0^t dt'[\bar{x}_2^2(t')-\bar{y}_2^2(t')]
\end{aligned} \quad (18)
$$

The expression in Eq. (18) can be transformed in to

$$
\begin{aligned}
J(x_1, x_2, y_1, y_2, t; x_1', x_2', y_1', y_2', 0) = \int \mathsf{D}\bar{x}_1 \mathsf{D}\bar{y}_1 \mathsf{D}\bar{x}_2 \mathsf{D}\bar{y}_2 \exp\frac{i}{\hbar}\{S_{A1}[\bar{x}_1] - S_{A1}[\bar{y}_1] \\
+S_{A2}[\bar{x}_2] - S_{A2}[\bar{y}_2] + S_{12}[\bar{x}_1, \bar{x}_2] - S_{12}[\bar{y}_1, \bar{y}_2] + S_{ext}^{(1)}[\bar{x}_1] - S_{ext}^{(1)}[\bar{y}_1] + S_{ext}^{(2)}[\bar{x}_2] - S_{ext}^{(2)}[\bar{y}_2] \\
-M_1\gamma_1 \int_0^t dt'[\bar{x}_1(t')\dot{\bar{x}}_1(t') - \bar{y}_1(t')\dot{\bar{y}}_1(t') + \bar{x}_1(t')\dot{\bar{y}}_1(t') - \bar{y}_1(t')\dot{\bar{x}}_1(t')] \\
-M_2\gamma_2 \int_0^t dt'[\bar{x}_2(t')\dot{\bar{x}}_2(t') - \bar{y}_2(t')\dot{\bar{y}}_2(t') + \bar{x}_2(t')\dot{\bar{y}}_2(t') - \bar{y}_2(t')\dot{\bar{x}}_2(t')]\} \exp[-(\phi_1 + \phi_2)]
\end{aligned}, \quad (19)
$$

where we desigtated $\gamma_{1,2} = \eta_{1,2}/2M_{1,2}$, and $\phi_i \equiv \phi_i[\bar{x}_i, \bar{y}_i]$ are equal

$$
\phi_i = \frac{2M_i\gamma_i}{\hbar\pi}\int_0^{v_i^{\max}} d\omega\, \omega\, Coth\left(\frac{\hbar\omega}{2k_B T_i}\right)\int_0^t dt'\int_0^{t'} dt''\{\bar{x}_i(t')-\bar{y}_i(t')\}\cos[\omega(t'-t'')]\{\bar{x}_i(t'')-\bar{y}_i(t'')\}, \quad (20)
$$

where $i = 1, 2$.

The Eq. (19) for propagator can be rewritten in the compact form as follows

$$
J(x_1, x_2, y_1, y_2, t; x_1', x_2', y_1', y_2', 0) = \int \mathsf{D}\bar{x}_1 \mathsf{D}\bar{y}_1 \mathsf{D}\bar{x}_2 \mathsf{D}\bar{y}_2 \exp\frac{i}{\hbar}S[\bar{x}_1, \bar{x}_2, \bar{y}_1, \bar{y}_2]\exp[-(\phi_1 + \phi_2)], \quad (21)
$$

where

$$
S[\bar{x}_1, \bar{x}_2, \bar{y}_1, \bar{y}_2] = \int_0^t dt'\, \mathsf{L}(\bar{x}_1, \bar{x}_2, \dot{\bar{x}}_1, \dot{\bar{x}}_2, \bar{y}_1, \bar{y}_2, \dot{\bar{y}}_1, \dot{\bar{y}}_2). \quad (22)
$$

Then, we write the Lagrangian in Eq. (22) omitting the straight lines over the letters for brevity

$$
\begin{aligned}
\mathsf{L} = M_1\dot{x}_1^2/2 - M_1\dot{y}_1^2/2 + M_2\dot{x}_2^2/2 - M_2\dot{y}_2^2/2 \\
-M_1\omega_{01}^2 x_1^2/2 + M_1\omega_{01}^2 y_1^2/2 - M_2\omega_{02}^2 x_2^2/2 + M_2\omega_{02}^2 y_2^2/2 \\
-M_1\gamma_1[x_1\dot{x}_1 - y_1\dot{y}_1 + x_1\dot{y}_1 - y_1\dot{x}_1] - M_2\gamma_2[x_2\dot{x}_2 - y_2\dot{y}_2 + x_2\dot{y}_2 - y_2\dot{x}_2] \\
+\lambda x_1 x_2 - \lambda y_1 y_2 + x_1 f_1(\tau) - y_1 f_1(\tau) + x_2 f_2(\tau) - y_2 f_2(\tau)
\end{aligned}. \quad (23)
$$

The Lagrange equations of motion

$$
\frac{d}{dt}\frac{\partial \mathsf{L}}{\partial \dot{x}_i} - \frac{\partial \mathsf{L}}{\partial x_i} = 0, \quad \frac{d}{dt}\frac{\partial \mathsf{L}}{\partial \dot{y}_i} - \frac{\partial \mathsf{L}}{\partial y_i} = 0, \quad (i = 1, 2), \quad (24)
$$

as applied to the Lagrangian in Eq. (23) give the following system of coupled equations for driven oscillators

$$\begin{cases} \ddot{x}_1 + 2\gamma_1 \dot{y}_1 + \omega_{01}^2 x_1 = (\lambda/M_1)x_2 + f_1(\tau)/M_1 \\ \ddot{x}_2 + 2\gamma_2 \dot{y}_2 + \omega_{02}^2 x_2 = (\lambda/M_2)x_1 + f_2(\tau)/M_2 \\ \ddot{y}_1 + 2\gamma_1 \dot{x}_1 + \omega_{01}^2 y_1 = (\lambda/M_1)y_2 - f_1(\tau)/M_1 \\ \ddot{y}_2 + 2\gamma_2 \dot{x}_2 + \omega_{02}^2 y_2 = (\lambda/M_2)y_1 - f_2(\tau)/M_2 \end{cases}. \tag{25}$$

Solution of the system is done in Appendix A. The classical solution is obtained in new variables $X_{1,2} = x_{1,2} + y_{1,2}$, $\xi_{1,2} = x_{1,2} - y_{1,2}$. In these new variables the Lagrangian in Eq. (23) reads

$$\begin{aligned} \mathsf{L} = & M_1 \dot{X}_1 \dot{\xi}_1 / 2 - M_1 \omega_{01}^2 X_1 \xi_1 / 2 - M_1 \gamma_1 \dot{X}_1 \xi_1 \\ & + M_2 \dot{X}_2 \dot{\xi}_2 / 2 - M_2 \omega_{02}^2 X_2 \xi_2 / 2 - M_2 \gamma_2 \dot{X}_2 \xi_2 \\ & + (\lambda/2)(X_1 \xi_2 + X_2 \xi_1) + \xi_1 f_1(\tau) + \xi_2 f_2(\tau) \end{aligned}. \tag{26}$$

Now, we represent the paths expressed in new variables as the sums $X_{1,2} = \tilde{X}_{1,2} + X'_{1,2}$, $\xi_{1,2} = \tilde{\xi}_{1,2} + \xi'_{1,2}$, explicitly selecting classical paths in Eqs.(A25, A26)-(A28, A29) and fluctuating parts $X'_{1,2}$, $\xi'_{1,2}$ with boundary conditions $X'_{1,2}(0) = X'_{1,2}(t) = 0$, $\xi'_{1,2}(0) = \xi'_{1,2}(t) = 0$. The Lagrangian in Eq.(26) becomes

$$\begin{aligned} \mathsf{L} \equiv \tilde{\mathsf{L}} + \mathsf{L}' = & M_1 \dot{\tilde{X}}_1 \dot{\tilde{\xi}}_1 / 2 - M_1 \omega_{01}^2 \tilde{X}_1 \tilde{\xi}_1 / 2 - M_1 \gamma_1 \dot{\tilde{X}}_1 \tilde{\xi}_1 \\ & + M_2 \dot{\tilde{X}}_2 \dot{\tilde{\xi}}_2 / 2 - M_2 \omega_{02}^2 \tilde{X}_2 \tilde{\xi}_2 / 2 - M_2 \gamma_2 \dot{\tilde{X}}_2 \tilde{\xi}_2 + (\lambda/2)(\tilde{X}_1 \tilde{\xi}_2 + \tilde{X}_2 \tilde{\xi}_1) \\ & + \tilde{\xi}_1 f_1(\tau) + \tilde{\xi}_2 f_2(\tau) \\ & + M_1 \dot{X}'_1 \dot{\xi}'_1 / 2 - M_1 \omega_{01}^2 X'_1 \xi'_1 / 2 - M_1 \gamma_1 \dot{X}'_1 \xi'_1 \\ & + M_2 \dot{X}'_2 \dot{\xi}'_2 / 2 - M_2 \omega_{02}^2 X'_2 \xi'_2 / 2 - M_2 \gamma_2 \dot{X}'_2 \xi'_2 + (\lambda/2)(X'_1 \xi'_2 + X'_2 \xi'_1) \\ & + \xi'_1 f_1(\tau) + \xi'_2 f_2(\tau) \end{aligned}. \tag{27}$$

With the above described assumptions and taking into account Eq.(27) the action in Eq.(22) calculated in new variables $S = \tilde{S}_{cl} + S'$ is expressed as follows

$$\begin{aligned} \tilde{S}_{cl} &= \int_0^t \tilde{\mathsf{L}}(\tau) d\tau = \tilde{S}_{cl}^{(1)} + \tilde{S}_{cl}^{(2)} + \tilde{S}_{cl}^{(12)} + \tilde{S}_{ext}^{(1)} + \tilde{S}_{ext}^{(2)}, \\ S' &= \int_0^t \mathsf{L}'(\tau) d\tau = S'_1 + S'_2 + S'_{12} + S'^{(1)}_{ext} + S'^{(2)}_{ext} \end{aligned}, \tag{28}$$

where

$$\begin{aligned} \tilde{S}_{cl}^{(1)} + \tilde{S}_{cl}^{(2)} + \tilde{S}_{ext}^{(1)} + \tilde{S}_{ext}^{(2)} = & D_1(X_{f1}\xi_{f1}) + [D_5 + D'_5](X_{f2}\xi_{f1}) + [D_6 + D'_6](X_{f1}\xi_{f2}) \\ & + D'_1(X_{f2}\xi_{f2}) + D_2(X_{i1}\xi_{f1}) + D_3(X_{f1}\xi_{i1}) + D_4(X_{i1}\xi_{i1}) + D'_2(X_{i2}\xi_{f2}) + D'_3(X_{f2}\xi_{i2}) \\ & + D'_4(X_{i2}\xi_{i2}) + [D_7 + D'_7](X_{i1}\xi_{f2}) + [D_8 + D'_8](X_{i2}\xi_{f1}) + [D_9 + D'_9](X_{f1}\xi_{i2}) \\ & + [D_{10} + D'_{10}](X_{f2}\xi_{i1}) + [D_{11} + D'_{11}](X_{i1}\xi_{i2}) + [D_{12} + D'_{12}](X_{i2}\xi_{i1}) \\ & + \int_0^t d\tau \, \xi_1(\tau) f_1(\tau) + \int_0^t d\tau \, \xi_2(\tau) f_2(\tau) \end{aligned}. \tag{29}$$

Then, we substitute into the inhomogeneous part the total classical trajectories for $\xi_{1,2}$ and separate constants and variables of integraion

$$\int_0^t d\tau\, \xi_1(\tau) f_1(\tau) + \int_0^t d\tau\, \xi_2(\tau) f_2(\tau) = \\ \int_0^t d\tau\, \xi_{p1}(\tau) f_1(\tau) + \int_0^t d\tau\, \xi_{p2}(\tau) f_2(\tau) + \Phi(t) + \Lambda_1(t)\xi_{i1} + \Lambda_2(t)\xi_{i2} \quad , \tag{30}$$

where $\Phi(t)$, $\Lambda_1(t)$, $\Lambda_2(t)$ are written in Appendix C.

Then we have taken into account the interactions in $\tilde{S}_{cl}^{(12)}$ between two oscillators

$$\begin{aligned}
\tilde{S}_{cl}^{(12)} = & \Pi_1(X_{f1}\xi_{f1}) + \Pi_2(X_{f1}\xi_{f2}) + \Pi_3(X_{f2}\xi_{f1}) + \Pi_4(X_{f2}\xi_{f2}) + \\
& + \Pi_5(X_{f1}\xi_{i1}) + \Pi_6(X_{f1}\xi_{i2}) + \Pi_7(X_{f2}\xi_{i1}) + \Pi_8(X_{f2}\xi_{i2}) + \\
& + \Pi_9(X_{i1}\xi_{f1}) + \Pi_{10}(X_{i1}\xi_{f2}) + \Pi_{11}(X_{i2}\xi_{f1}) + \Pi_{12}(X_{i2}\xi_{f2}) + \\
& + \Pi_{13}(X_{i1}\xi_{i1}) + \Pi_{14}(X_{i1}\xi_{i2}) + \Pi_{15}(X_{i2}\xi_{i1}) + \Pi_{16}(X_{i2}\xi_{i2})
\end{aligned} \quad , \tag{31}$$

with temporal functions $\Pi_k \equiv \Pi_k(t)$, $k=1,...,16$ from [45].

Finally, we have for the classical actions in Eq.(28)

$$\begin{aligned}
\tilde{S}_{cl}^{(1)} + \tilde{S}_{cl}^{(2)} + \tilde{S}_{cl}^{(12)} + \tilde{S}_{ext}^{(1)} + \tilde{S}_{ext}^{(2)} = & [D_1 + \Pi_1](X_{f1}\xi_{f1}) + [D_5 + D_5' + \Pi_3](X_{f2}\xi_{f1}) \\
& + [D_6 + D_6' + \Pi_2](X_{f1}\xi_{f2}) + [D_1' + \Pi_4](X_{f2}\xi_{f2}) + \{(D_2 + \Pi_9)\xi_{f1} \\
& + (D_7 + D_7' + \Pi_{10})\xi_{f2} + U_1(t)\}X_{i1} + \{(D_2' + \Pi_{12})\xi_{f2} + (D_8 + D_8' + \Pi_{11})\xi_{f1} + U_2(t)\}X_{i2} \\
& + \{(D_3 + \Pi_5)X_{f1} + (D_{10} + D_{10}' + \Pi_7)X_{f2} + \Lambda_1(t)\}\xi_{i1} + \{(D_3' + \Pi_8)X_{f2} \\
& + (D_9 + D_9' + \Pi_6)X_{f1} + \Lambda_2(t)\}\xi_{i2} + [D_4 + \Pi_{13}](X_{i1}\xi_{i1}) + [D_{11} + D_{11}' + \Pi_{14}](X_{i1}\xi_{i2}) \\
& + [D_{12} + D_{12}' + \Pi_{15}](X_{i2}\xi_{i1}) + [D_4' + \Pi_{16}](X_{i2}\xi_{i2}) + \int_0^t d\tau\, \xi_{p1}(\tau) f_1(\tau) + \int_0^t d\tau\, \xi_{p2}(\tau) f_2(\tau) \\
& + \Phi(t) + U_{01}(t)
\end{aligned} \tag{32}$$

where the terms $U_{01}(t), U_1(t), U_2(t),$ take into account inhomogeneous additional terms in actions

$$\begin{aligned}
[\tilde{S}_{cl}^{(1)}(t) + \tilde{S}_{cl}^{(2)}(t) + \tilde{S}_{cl}^{(12)}(t)]_{partial} = & \int_0^t d\tau \{M_1 \dot{\tilde{X}}_1 \dot{\xi}_{p1}/2 - M_1 \omega_1^2 \tilde{X}_1 \xi_{p1}/2 - M_1 \gamma_1 \dot{\tilde{X}}_1 \xi_{p1} \\
& + M_2 \dot{\tilde{X}}_2 \dot{\xi}_{p2}/2 - M_2 \omega_2^2 \tilde{X}_2 \xi_{p2}/2 - M_2 \gamma_2 \dot{\tilde{X}}_2 \xi_{p2} + (\lambda/2)[\tilde{X}_1 \xi_{p2} + \tilde{X}_2 \xi_{p1}]\} = \\
= & U_{01}(t) + U_1(t) X_{i1} + U_2(t) X_{i2}.
\end{aligned} \tag{33}$$

Then we should calculate of the functions $\phi_{1,2}$ in Eq.(21) taking into acoount the form of the classical solution with external forces

$$\begin{aligned}
\phi_i = & \frac{2M_i \gamma_i}{\hbar \pi} \int_0^{\nu_i^{max}} d\omega\, \omega\, Coth\left(\frac{\hbar\omega}{2k_B T_i}\right) \int_0^t dt' \int_0^{t'} dt'' \{\bar{x}_i(t') - \bar{y}_i(t')\} \cos[\omega(t'-t'')]\{\bar{x}_i(t'') - \bar{y}_i(t'')\} = \\
& \frac{2M_i \gamma_i}{\hbar \pi} \int_0^{\nu_i^{max}} d\omega\, \omega\, Coth\left(\frac{\hbar\omega}{2k_B T_i}\right) \int_0^t dt' \int_0^{t'} dt'' \{\bar{\xi}_i(t')\} \cos[\omega(t'-t'')]\{\bar{\xi}_i(t'')\} + \\
& \frac{2M_i \gamma_i}{\hbar \pi} \int_0^{\nu_i^{max}} d\omega\, \omega\, Coth\left(\frac{\hbar\omega}{2k_B T_i}\right) \int_0^t dt' \int_0^{t'} dt'' \{\xi_{pi}(t')\} \cos[\omega(t'-t'')]\{\xi_{pi}(t'')\}
\end{aligned} \quad , \tag{34}$$

where $i=1,2$.

The external forces are not lead to new results in Eq.(34) compare with previous calculations in [43-45], because related additional term in Eq.(34) has no variables for further integration. Calculations of a classical action and functions $\phi_{1,2}$ in Eq.(21) yield the final expression for the propagator in case of two driven coupled oscillators

$$J(X_{f1},X_{f2},\xi_{f1},\xi_{f2},t;X_{i1},X_{i2},\xi_{i1},\xi_{i2},0) = \tilde{C}_0(t)\exp\frac{i}{\hbar}\{\tilde{S}_{cl}^{(1)} + \tilde{S}_{cl}^{(2)} + \tilde{S}_{cl}^{(12)} + \tilde{S}_{ext}^{(1)} + \tilde{S}_{ext}^{(2)}\}$$

$$\times\exp-\frac{1}{\hbar}\{A_1(t)\bar{\xi}_{f1}^2 + B_1(t)\bar{\xi}_{f1}\bar{\xi}_{i1} + C_1(t)\bar{\xi}_{i1}^2\}\exp-\frac{1}{\hbar}\{A_2(t)\bar{\xi}_{f2}^2 + B_2(t)\bar{\xi}_{f2}\bar{\xi}_{i2} + C_2(t)\bar{\xi}_{i2}^2\}$$

$$\times\exp-\frac{1}{\hbar}\{E_1(t)\bar{\xi}_{i1}\bar{\xi}_{i2} + E_2(t)\bar{\xi}_{f2}\bar{\xi}_{i1} + E_3(t)\bar{\xi}_{f1}\bar{\xi}_{i2} + E_4(t)\bar{\xi}_{f1}\bar{\xi}_{f2}\}$$

$$\times G(X_{f1},X_{f2},\bar{\xi}_{f1},\bar{\xi}_{f2},t;X_{i1},X_{i2},\bar{\xi}_{i1},\bar{\xi}_{i2},0)$$

(35)

where all time dependent functions $A_{1,2}(t)$, $B_{1,2}(t)$, $C_{1,2}(t)$, $E_{1,2,3,4}(t)$ are written in [45]. The function $\tilde{C}_0(t)$ consists all irrelevant terms, which have no variables for further integration.

The fluctuational integral in Eq.(35) is

$$G(X_{f1},X_{f2},\xi_{f1},\xi_{f2},t;X_{i1},X_{i2},\xi_{i1},\xi_{i2},0) = \frac{1}{4}\int \mathsf{D}X_1'\mathsf{D}X_2'\mathsf{D}\xi_1'\mathsf{D}\xi_2'\exp\frac{i}{\hbar}\{S_1' + S_2' + S_{12}' + S_{ext}'\}$$

$$\times\exp-\frac{2}{\hbar}\{\phi_{T1}[\xi_1',\xi_1'] + \phi_{T2}[\xi_2',\xi_2']\}\exp-\frac{1}{\hbar}\{\phi_{T1}[\tilde{\xi}_1,\xi_1'] + \phi_{T2}[\tilde{\xi}_2,\xi_2']\}$$

(36)

where the integration is carried out along all closed paths because $X_{1,2}'(0) = X_{1,2}'(t) = 0$ and $\xi_{1,2}'(0) = \xi_{1,2}'(t) = 0$, and functions $\phi_{T1,T2}$ can be found in [43].

The fluctuational integral in Eq.(36) without of the additional term related to external forces is calculated in [43], and the additional term $S_{ext}'$ is not change our final result.

### B. Reduced density matrix of driven coupled oscillators

Thus, after calculating of the fluctuational integral the final form of the propagating function in Eq.(35) reads

$$J(X_{f1},X_{f2},\xi_{f1},\xi_{f2},t;X_{i1},X_{i2},\xi_{i1},\xi_{i2},0) = \tilde{C}(t)\exp\frac{i}{\hbar}\tilde{S}_{cl}^{(12)}$$

$$\times\exp\frac{i}{\hbar}\tilde{S}_{cl}^{(1)}\exp-\frac{1}{\hbar}\{A_1(t)\xi_{f1}^2 + B_1(t)\xi_{f1}\xi_{i1} + C_1(t)\xi_{i1}^2\}$$

$$\times\exp\frac{i}{\hbar}\tilde{S}_{cl}^{(2)}\exp-\frac{1}{\hbar}\{A_2(t)\xi_{f2}^2 + B_2(t)\xi_{f2}\xi_{i2} + C_2(t)\xi_{i2}^2\}$$

$$\times\exp-\frac{1}{\hbar}\{E_1(t)\xi_{i1}\xi_{i2} + E_2(t)\xi_{f2}\xi_{i1} + E_3(t)\xi_{f1}\xi_{i2} + E_4(t)\xi_{f1}\xi_{f2}\}$$

(37)

In order to calculate the reduced density matrix $\rho(x_1, x_2, y_1, y_2, t)$ in Eq.(12) we ought to assign the initial density matrixes $\rho_A^{(1)}(x_1', y_1', 0)$ and $\rho_A^{(2)}(x_2', y_2', 0)$ for two selected oscillators. In new variables we have instead of Eq.(12) the following expression for the reduced density matrix

$$\rho(X_{f1}, X_{f2}, \xi_{f1}, \xi_{f2}, t) = \int dX_{i1} dX_{i2} d\xi_{i1} d\xi_{i2}\, J(X_{f1}, X_{f2}, \xi_{f1}, \xi_{f2}, t; X_{i1}, X_{i2}, \xi_{i1}, \xi_{i2}, 0)$$
$$\times \rho_A^{(1)}(X_{i1}, \xi_{i1}, 0) \rho_A^{(2)}(X_{i2}, \xi_{i2}, 0)$$
(38)

where we choose the same initial states of oscillators as in [43-45]

$$\rho_A^{(k)}(X_{ik}, \xi_{ik}, 0) = (2\pi\sigma_{0k}^2)^{-1/2} \exp\left[-(X_{ik}^2 + \xi_{ik}^2)/8\sigma_{0k}^2\right], \quad (k=1,2)$$
(39)

where $\sigma_{0k}^2$, $(k=1,2)$ are the initial spatial variances of two oscillators.

Using the propagator from Eq.(37) and Eq.(39) an integration in Eq.(38) is straghtforward and leads to an explicit expression for the reduced density matrix of two driven coupled oscillators, which is valid for any time, including a steady state regime.

In case of external forces the total non-Hermitian, in general, density matrix is as follows

$$\rho(X_{f1}, X_{f2}, t; \xi_{f1}, \xi_{f2}, 0) = C(t) \exp\left\{-\left[g_1(t)X_{f1}^2 + g_{12}(t)X_{f1}X_{f2} + g_2(t)X_{f2}^2\right]\right\}$$
$$\times \exp\left\{-\left[g_1'(t)\xi_{f1}^2 + g_{12}'(t)\xi_{f1}\xi_{f2} + g_2'(t)\xi_{f2}^2\right]\right\}$$
$$\times \exp\left\{i\left[g_{11}''(t)X_{f1}\xi_{f1} + g_{21}''(t)X_{f2}\xi_{f1} + g_{12}''(t)X_{f1}\xi_{f2} + g_{22}''(t)X_{f2}\xi_{f2}\right]\right\}$$
$$\times \exp\left\{iA_{ext}^+(t)X_{f1} + iB_{ext}^+(t)X_{f2}\right\} \exp\left\{iA_{ext}^-(t)\xi_{f1} + iB_{ext}^-(t)\xi_{f2}\right\}$$
$$\times \exp\left\{-a_{ext}^+(t)X_{f1} - b_{ext}^+(t)X_{f2}\right\} \exp\left\{-a_{ext}^-(t)\xi_{f1} - b_{ext}^-(t)\xi_{f2}\right\},$$
(40)

where $C(t)$ is the normalization constant, $g(t)$, $g'(t)$, $g''(t)$, $A_{ext}^\pm(t)$, $B_{ext}^\pm(t)$, $a_{ext}^\pm(t)$, $b_{ext}^\pm(t)$ are pure real functions. The non-Hermiticity of the $\rho$ in Eq.(40) can be easily proved recalling that $X_f = x_f + y_f$, $\xi_f = x_f - y_f$ and taking into account the reality properties of the just above mentioned functions. The Hermitian part of the of the $\rho$ in Eq.(40) can be easily extracted representing $\rho$ as usual as a sum of the Hermitian and anti-Hermitian parts. But, it is occure in this problem that $|A^+| \ll |A^-|$, $|B^+| \ll |B^-|$ and $|a^+| \gg |a^-|$, $|b^+| \gg |b^-|$. That is why we can put

$$A_{ext}^+ X_{f1} + B_{ext}^+ X_{f2} + A_{ext}^- \xi_{f1} + B_{ext}^- \xi_{f2} \approx A_{ext}^- x_{f1} - A_{ext}^- y_{f1} + B_{ext}^- x_{f2} - B_{ext}^- y_{f2},$$
$$a_{ext}^+ X_{f1} + b_{ext}^+ X_{f2} + a_{ext}^- \xi_{f1} + b_{ext}^- \xi_{f2} \approx a_{ext}^+ x_{f1} + a_{ext}^+ y_{f1} + b_{ext}^+ x_{f2} + b_{ext}^+ y_{f2}$$
(41)

The Hermitian density matrix in this case is as follows

$$\rho(X_{f1}, X_{f2}, \xi_{f1}, \xi_{f2}, t) = C(t) \exp\left\{-\left[g_1(t)X_{f1}^2 + g_{12}(t)X_{f1}X_{f2} + g_2(t)X_{f2}^2\right]\right\}$$
$$\times \exp\left\{-\left[g_1'(t)\xi_{f1}^2 + g_{12}'(t)\xi_{f1}\xi_{f2} + g_2'(t)\xi_{f2}^2\right]\right\}$$
$$\times \exp\left\{i\left[g_{11}''(t)X_{f1}\xi_{f1} + g_{21}''(t)X_{f2}\xi_{f1} + g_{12}''(t)X_{f1}\xi_{f2} + g_{22}''(t)X_{f2}\xi_{f2}\right]\right\}$$
$$\times \exp\left\{-a_{ext}^+(t)X_{f1} - b_{ext}^+(t)X_{f2}\right\} \exp\left\{iA_{ext}^-(t)\xi_{f1} + iB_{ext}^-(t)\xi_{f2}\right\}$$
(42)

where $g(t)$, $g'(t)$, $g''(t)$ and $a_{ext}^+(t)$, $b_{ext}^+(t)$, $A_{ext}^-(t)$, $B_{ext}^-(t)$ are written down in Appendix B.

The Hermiticity $\rho(x_{f1}, x_{f2}, y_{f1}, y_{f2}, t) = \rho^*(y_{f1}, y_{f2}, x_{f1}, x_{f2}, t)$ of the matrix in Eq.(42) is clear now and all mean values of observables are real, see the Appendix C.

Also, we can to represent the non-Hermitian matrix in Eq.(40) as follows

$$\rho(X_{f1}, X_{f2}, t; \xi_{f1}, \xi_{f2}, 0) = C(t)\rho_H(t)\rho_N(t) , \qquad (43)$$

where the Hermitian part $\rho_H(t)$ is the matrix from Eq.(42) and the non-Hermitian part

$$\rho_N(t) = \exp\left\{iA_{ext}^+(t)X_{f1} + iB_{ext}^+(t)X_{f2}\right\}\exp\left\{-a_{ext}^-(t)\xi_{f1} - b_{ext}^-(t)\xi_{f2}\right\} . \qquad (44)$$

Then, representing the total matrix as a sum of the Hermitian and anti-Hermitian parts we can to obtain

$$\rho = C[\rho_H(\rho_N + \rho_N^+)/2 + \rho_H(\rho_N - \rho_N^+)/2] \approx C\rho_H , \qquad (45)$$

which can be fullfilled under restrictions $A_{ext}^+ = -A_{ext}^+ = 0$, $B_{ext}^+ = -B_{ext}^+ = 0$ and $a_{ext}^- = -a_{ext}^- = 0$, $b_{ext}^- = -b_{ext}^- = 0$. These relations are in agreement with approximations accepted in Eq.(41).

### C. Temporal dynamics of mean values

All elements of a covariance matrix for the driven coupled oscillators are obtained in Appendix C using the Hermitian density matrix from Eq.(42). It should be noted that all dispersions of coordinates and momenta in this problem are identical to the problem of coupled quantum oscillators without external forces. But, the mean quadratic values of coordinates and momenta are essetially different in these problems. Besides, mean values of coordinates and momenta are not equal to zero in the problem under our study, contrary to the problem without of external forces, where the mean values of coordinates and momenta are equal zero. We investigated dispersions in our prevoius papers in [43-45] that is why we put our main attention to study the temporal dynamics of mean values of coordinates of coupled oscillators under the action of external forces.

For our study we chose the exponential-like external forces

$$f_i(\tau) = f_{0i}\theta(\tau - t_{0i})Exp(-\delta_i\tau), \quad (i = 1, 2), \qquad (46)$$

which are switched on at different times $t_{0i}$, $(i = 1, 2)$ in general, where $f_{0i}$ is the force amplitude, $\theta$ is the unit step function, $\delta_i$ is the decaying factor.

To chose the appropriate amplitudes of the forces we consider that the total increase of momenta of an oscillator during a time interval from $\tau = 0$ to $\tau = t$ is as follows

$$\Delta p_i = p_i = \int_0^t d\tau\, f(\tau) \simeq f_{0i}\delta_i^{-1} \exp(-\delta_i t_{0i}). \qquad (47)$$

Then, we use the virial theorem to obtain $f_{0i} \approx \delta_i \exp(\delta_i t_{0i})\sqrt{<x^2>}$, where $<x^2>$, for example, can be of the dispersion $<x^2> = \sigma_{0i}^2 = \hbar/2M_i \omega_{0i}$.

For numerical calculations we have chosen the following parameters of oscillators:

$M_1 = 10^{-23} g$, $M_2 = 5M_1$, $\omega_{01} = 10^{13} rad/s$, $\omega_{02} = 3\omega_{01}$, $\gamma_1 = \gamma_2 = \gamma = 0.01\omega_{01}$, relating to solid materials, and $\delta_1 = 10\gamma$, $\delta_2 = 10\gamma$, $t_{01} = 10^{-13} s$, $t_{02} = 10^{-12} s$ for the external forces.

Fig.2 exemplifies the normalized mean values of the coordinates $\bar{x}_i(t)/\sqrt{\sigma_{0i}^2}$, $(i=1,2)$ of the first –a) and second –b) oscillators versus a time $t$, where $\sigma_{0i}^2 = \hbar/2M_i\omega_{0i}$, $(i=1,2)$ using Eq.(C10, C18). These figures correspond to the case of no coupling $\lambda = 0$, and to the case of the total equilibrium, when $T_1 = T_2 = 300K$. The profiles of external forces $f_{1,2}(t)$ of unit amplitude and of different signs in accordance with Eq.(46) is shown between of these graphs, as well as in figures 3,4. It is clearly seen that the mean values tends to zero, when both of the forces $f_{1,2} \to 0$. Also, we can see absolutely independent and different dynamics of selected oscillators, as it must be at $\lambda = 0$ and at different parameters of oscillators and forces.

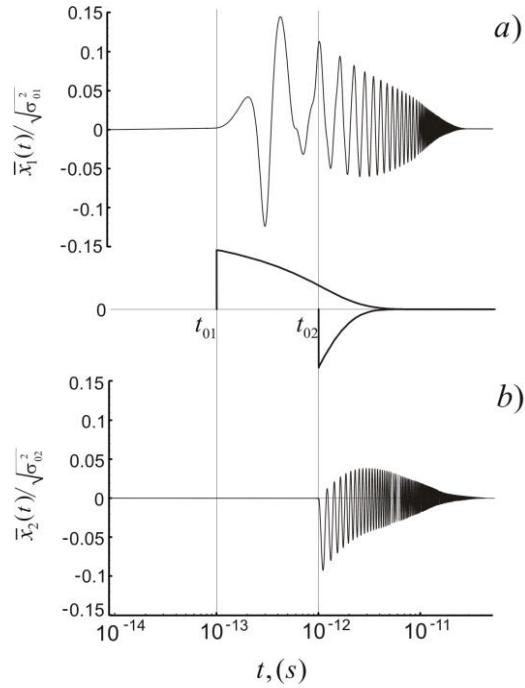

Fig.2

Fig.3 shows the normalized mean values of the coordinates $\bar{x}_i(t)/\sqrt{\sigma_{0i}^2}$, $(i=1,2)$ of the first –a) and second –b) oscillators versus a time $t$ using Eq.(C10, C18) at relatively strong coupling $\tilde{\lambda} = \lambda / \omega_{01}\omega_{02}\sqrt{M_1 M_2} = 0.3$ between of oscillators. These figures also correspond to the case of the total equilibrium, when $T_1 = T_2 = 300K$. As well as in the previous figure it is clearly seen that the mean values tends to zero, when both of the forces $f_{1,2} \to 0$. But, in this case we can see that the first external force also acts on the second oscillator due to coupling at $t_{01} < t < t_{02}$ when $f_2 = 0$. This value of the coupling constant has been chosen because of the model of bilinear coupling breaks down at stronger couplings, see, for instance, [44, 45].

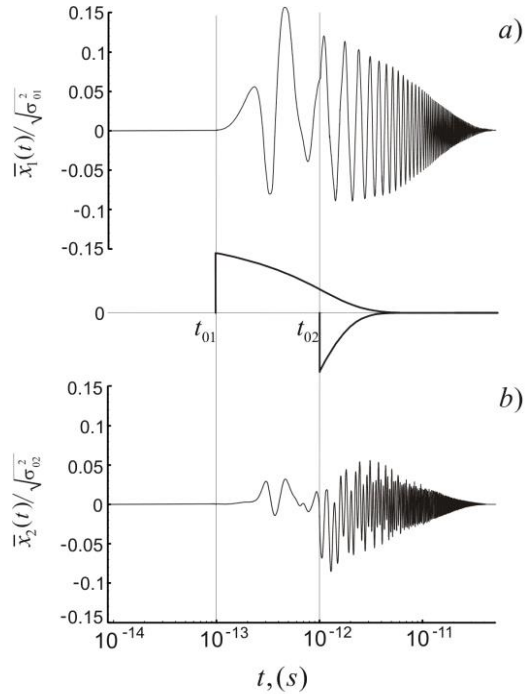

Fig.3

The normalized mean values of the coordinates $\bar{x}_i(t)/\sqrt{\sigma_{0i}^2}$, $(i = 1, 2)$ of the first –a) and second – b) oscillators versus a time $t$ using Eq.(C10, C18) at $\tilde{\lambda} = 0.3$ between of oscillators are shown in Fig.4 in the case out of total equilibrium in the system, when $T_1 = 300K$ and $T_2 = 900K$. A comparison of figures 3 and 4 shows that the difference in temperatures yields in a different dynamics of coupled oscillators. It should be noted that despite of parameters and thermodynamical conditions in the problem under study, the mean values of coordinates tends to zero at the force zeroing. This corresponds to $A_{ext}^-(t) \to 0$, $B_{ext}^-(t) \to 0$, $a_{ext}^+(t) \to 0$, $b_{ext}^+(t) \to 0$ of the density matrix in Eq.(42). The functions $A_{ext}^-(t)$, $B_{ext}^-(t)$, $a_{ext}^+(t)$, $b_{ext}^+(t)$ in Eq.(42) are not zero only in the case when the external forces are not zero, see corresponding formulas in Appendix B.

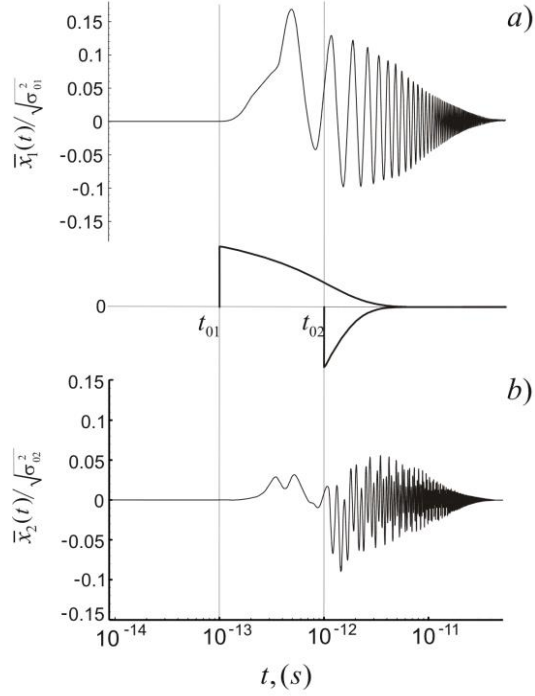

Fig.4

It should be noted that in calculations of mean values we can use arbitrary forms of external forces including single impulses, which are finite in time, or impulse sequences with given on-off time ratio, steady state forces alternating in time, constant in time and so on, which can be applied at any time $t \geq 0$ to any of two coupled oscillators. Besides, the selected oscillators can be characterized by arbitrary properties at different temperatures of separated baths, in general.

## IV. Conclusion

Our paper is devoted to study of the relaxation problem of open quantum systems driven by external forces. We considered two bilinear coupled oscillators and, in its turn, each of these oscillators is coupled with separate reservoirs of harmonic oscillators and subjected by external forces. Corresponding Hamiltonian is provided. In the initial time all interactions among oscillators are switched on and maintained during arbitrary time interval. Then, the arbitrary external forces begin acting at arbitrary time moments. Using the path integral methods we found and analyzed an analytical expression for time-dependent density matrix of two forced coupled quantum oscillators interacting with different reservoirs of oscillators. We calculated corresponding propagator in this case. All elements of the covariance matrix are calculated using the known reduction procedure. It is shown that the mean values of coordinates and momenta of coupled oscillators are not zero in case of externally driven oscillators. Time-dependent behavior of the mean values at different conditions is graphically illustrated. Coupled dynamics of selected oscillators at relatively large coupling constants is demonstrated at different thermodynamic conditions. It is interesting to note that the mean quadratic characteristics of oscillators are different for the case of "freely developing" pair of oscillators and for the driven pair of oscillators, but their dispersions are identically equal.

**Acknowledgement**

**Appendix A. Solution to the system in Eq.(25) for classical paths with external forces.**

The total analysis of the coupled motion has been done on the basis of the textbooks [53-55]. In new variables we obtained from Eq.(25) two pairs of coupled equations for classical paths

$$\begin{cases} \ddot{\tilde{X}}_1 + 2\gamma_1 \dot{\tilde{X}}_1 + \omega_{01}^2 \tilde{X}_1 - (\lambda/M_1)\tilde{X}_2 = 0 \\ \ddot{\tilde{X}}_2 + 2\gamma_2 \dot{\tilde{X}}_2 + \omega_{02}^2 \tilde{X}_2 - (\lambda/M_2)\tilde{X}_1 = 0 \end{cases}, \quad (A1)$$

$$\begin{cases} \ddot{\tilde{\xi}}_1 - 2\gamma_1 \dot{\tilde{\xi}}_1 + \omega_{01}^2 \tilde{\xi}_1 - (\lambda/M_1)\tilde{\xi}_2 = 2f_1(\tau)/M_1 \\ \ddot{\tilde{\xi}}_2 - 2\gamma_2 \dot{\tilde{\xi}}_2 + \omega_{02}^2 \tilde{\xi}_2 - (\lambda/M_2)\tilde{\xi}_1 = 2f_2(\tau)/M_2 \end{cases}. \quad (A2)$$

A general solution of the first pair of homogeneous equations can be found in [43]. In its turn, the second pair of equations in (A2) for the classical paths $\tilde{\xi}_{1,2}(\tau)$ of backward amplitudes must be solved taking into account partial solutions $\xi_{p1}(\tau)$ and $\xi_{p2}(\tau)$ of inhomogeneous equations

$$\begin{aligned} \tilde{\xi}_1(\tau) &= B_1 \sin(\Omega_1 \tau + \varphi_1)\exp(-\delta_1 \tau) + r_2 B_2 \sin(\Omega_2 \tau + \varphi_2)\exp(-\delta_2 \tau) + \xi_{p1}(\tau), \\ \tilde{\xi}_2(\tau) &= r_1 B_1 \sin(\Omega_1 \tau + \varphi_1)\exp(-\delta_1 \tau) + B_2 \sin(\Omega_2 \tau + \varphi_2)\exp(-\delta_2 \tau) + \xi_{p2}(\tau), \end{aligned} \quad (A3)$$

where

$$B_1 \sin \varphi_1 = \frac{(\xi_{i1} - \xi_{p1}^0) - r_2(\xi_{i2} - \xi_{p2}^0)}{(1 - r_1 r_2)}, \quad B_2 \sin \varphi_2 = \frac{(\xi_{i2} - \xi_{p2}^0) - r_1(\xi_{i1} - \xi_{p1}^0)}{(1 - r_1 r_2)}, \quad (A4)$$

$$\begin{aligned} B_1 \cos \varphi_1 &= \frac{(\xi_{f1} - \xi_{p1}^t) - r_2(\xi_{f2} - \xi_{p2}^t)}{(1 - r_1 r_2)\sin(\Omega_1 t)}\exp(-\delta_1 t) - \cot(\Omega_1 t)\frac{(\xi_{i1} - \xi_{p1}^0) - r_2(\xi_{i2} - \xi_{p2}^0)}{(1 - r_1 r_2)}, \\ B_2 \cos \varphi_2 &= \frac{(\xi_{f2} - \xi_{p2}^t) - r_2(\xi_{f1} - \xi_{p1}^t)}{(1 - r_1 r_2)\sin(\Omega_2 t)}\exp(-\delta_2 t) - \cot(\Omega_2 t)\frac{(\xi_{i2} - \xi_{p2}^0) - r_1(\xi_{i1} - \xi_{p1}^0)}{(1 - r_1 r_2)}, \end{aligned} \quad (A5)$$

where we use the brief notations $\xi_{pk}(\tau = 0) \equiv \xi_{pk}^0$ and $\xi_{pk}(\tau = t) \equiv \xi_{pk}^t$, $(k = 1, 2)$ for the partial solutions.

It should be noted that the partial solution can be chosen that $\xi_{p1}^0 = \xi_{p2}^0 = 0$, $\xi_{p1}^t = \xi_{p2}^t = 0$. Corresponding way to obtain such a solution is described just below.

Let's designate in (A2) $F_1(\tau) = 2f_1(\tau)/M_1$, $F_2(\tau) = 2f_2(\tau)/M_2$ and represent the right part in (A2) as follows

$$F_k(\tau) = \text{Re}\{\int_{-\infty}^{\infty} F_k(\omega)\exp(i\omega\tau)d\omega/2\pi\}, \quad (k=1,2) \tag{A6}$$

We seek for the partial solution of the system (A2) in the same form

$$\xi_{pk}(\tau) = \text{Re}\{\int_{-\infty}^{\infty} g_k(\omega)\exp(i\omega\tau)d\omega/2\pi\}, \quad (k=1,2) \tag{A7}$$

Substitution (A6),(A7) into the system in (A2) yields in

$$g_1(\omega) = \frac{(\lambda/M_1)f_2(\omega) + D_2(\omega)f_1(\omega)}{D_1(\omega)D_2(\omega) - \lambda^2/M_1M_2} + A_1(\omega)\delta[D(\omega)], \tag{A8}$$

$$g_2(\omega) = \frac{(\lambda/M_2)f_1(\omega) + D_1(\omega)f_2(\omega)}{D_1(\omega)D_2(\omega) - \lambda^2/M_1M_2} + A_2(\omega)\delta[D(\omega)], \tag{A9}$$

where $D(\omega) = D_1(\omega)D_2(\omega) - \lambda^2/M_1M_2$, $D_k(\omega) = \omega_{0k}^2 - \omega^2 - i(2\gamma_k\omega)$, $(k=1,2)$, $\delta[D]$ is the Dirac delta function, $A_{1,2}(\omega)$ are unknown functions, which properties can be found from the analyticity of $g_{1,2}(\omega)$ and from boundary conditions for $\xi_{p1,p2}(\tau)$ at $\tau = 0, \tau = t$. The delta-functional terms in (A8),(A9) appear in the solutions due to reasoning described, for instance in [56].

From Eqs.(A8),(A9) we can see that the vibrations of coupled oscillators are determined by both of external forces via their coupling constant.

The determinant equation $D(\omega) = D_1(\omega)D_1(\omega) - \lambda^2/M_1M_2 = 0$ has different in general simple poles as the roots of the equation $\omega^4 + ia\omega^3 + b\omega^2 + ic\omega + d = 0$ where $a,b,c,d$ are pure real numbers

$$\begin{aligned} a &= 2(\gamma_1 + \gamma_1), \quad b = -(\omega_{01}^2 + \omega_{02}^2 + 4\gamma_1\gamma_2), \\ c &= -2(\gamma_2\omega_{01}^2 + \gamma_1\omega_{02}^2), \quad d = \omega_{01}^2\omega_{02}^2 - \lambda^2/M_1M_2. \end{aligned} \tag{A10}$$

The above equation of the fourth order has four different complex roots $\omega_{1,4} = \mp\tilde{\omega}_1 - i\gamma$ and $\omega_{2,3} = \mp\tilde{\omega}_2 - i\gamma$. Further, we no need the cumbersome explicit expressions for $\tilde{\omega}_{1,2}$.

Taking into account (A7)-(A9) and properties of the Kronecker delta-function we obtain

$$\xi_{p1,p2}(\tau) = \text{Re}\int_{-\infty}^{\infty}\frac{d\omega}{2\pi}C_{1,2}(\omega)e^{i\omega\tau} + \frac{1}{2\pi}\text{Re}\left\{\sum_{j=1}^{4}\frac{A_{1,2}(\omega_j)e^{i\omega_j\tau}}{|D'(\omega_j)|}\right\}, \tag{A11}$$

where $D'(\omega_j)$ is the first derivation of the determinant function at the roots $\omega_j$, $(j=1,...,4)$, and

$$C_1(\omega) = \frac{(\lambda/M_1)F_2(\omega) + D_2(\omega)F_1(\omega)}{D_1(\omega)D_2(\omega) - \lambda^2/M_1M_2}, \quad C_2(\omega) = \frac{(\lambda/M_2)F_1(\omega) + D_1(\omega)F_2(\omega)}{D_1(\omega)D_2(\omega) - \lambda^2/M_1M_2}. \tag{A12}$$

It is clear from the second parts in (A11) that the sums must be pure real. This allows putting some useful restrictions on the unknown complex functions $A_{1,2}(\omega)$. Taking into account that $|D'(\omega_1)|=|D'(\omega_4)|$ and $|D'(\omega_2)|=|D'(\omega_3)|$ because $\omega_{1,4}=\mp\tilde{\omega}_1-i\gamma$ and $\omega_{2,3}=\mp\tilde{\omega}_2-i\gamma$, and representing the unknown functions as $A_{1,2}=A'_{1,2}+iA''_{1,2}$ we have

$$\mathrm{Re}\sum_{j=1}^{4}\frac{A_{1,2}(\omega_j)\mathrm{e}^{i\omega_j\tau}}{|D'(\omega_j)|}=\frac{\mathrm{e}^{\gamma\tau}}{|D'(\omega_1)|}\mathrm{Re}\left[A_{1,2}(\omega_1)\mathrm{e}^{-i\tilde{\omega}_1\tau}+A_{1,2}(\omega_4)\mathrm{e}^{i\tilde{\omega}_1\tau}\right]$$
$$+\frac{\mathrm{e}^{\gamma\tau}}{|D'(\omega_2)|}\mathrm{Re}\left[A_{1,2}(\omega_2)\mathrm{e}^{-i\tilde{\omega}_2\tau}+A_{1,2}(\omega_3)\mathrm{e}^{i\tilde{\omega}_2\tau}\right],$$
(A13)

where

$$\mathrm{Re}\{A_{1,2}(\omega_1)\mathrm{e}^{-i\tilde{\omega}_1\tau}+A_{1,2}(\omega_4)\mathrm{e}^{i\tilde{\omega}_1\tau}\}+i\,\mathrm{Im}\{A_{1,2}(\omega_1)\mathrm{e}^{-i\tilde{\omega}_1\tau}+A_{1,2}(\omega_4)\mathrm{e}^{i\tilde{\omega}_1\tau}\}$$
$$=[A'_{1,2}(\omega_1)+A'_{1,2}(\omega_4)]\cos\tilde{\omega}_1\tau+[A''_{1,2}(\omega_1)-A''_{1,2}(\omega_4)]\sin\tilde{\omega}_1\tau$$
$$+i\{[A''_{1,2}(\omega_1)+A''_{1,2}(\omega_4)]\cos\tilde{\omega}_1\tau+[A'_{1,2}(\omega_4)-A'_{1,2}(\omega_1)]\sin\tilde{\omega}_1\tau\}$$
(A14)

and

$$\mathrm{Re}\{A_{1,2}(\omega_1)\mathrm{e}^{-i\tilde{\omega}_2\tau}+A_{1,2}(\omega_4)\mathrm{e}^{i\tilde{\omega}_2\tau}\}+i\,\mathrm{Im}\{A_{1,2}(\omega_1)\mathrm{e}^{-i\tilde{\omega}_2\tau}+A_{1,2}(\omega_4)\mathrm{e}^{i\tilde{\omega}_2\tau}\}$$
$$=[A'_{1,2}(\omega_2)+A'_{1,2}(\omega_3)]\cos\tilde{\omega}_2\tau+[A''_{1,2}(\omega_2)-A''_{1,2}(\omega_3)]\sin\tilde{\omega}_2\tau$$
$$+i\{[A''_{1,2}(\omega_2)+A''_{1,2}(\omega_3)]\cos\tilde{\omega}_2\tau+[A'_{1,2}(\omega_3)-A'_{1,2}(\omega_2)]\sin\tilde{\omega}_2\tau\},$$
(A15)

Because in Eq.(A14),(A15) we need only the pure real parts, the pure imaginary parts in these equations can be subjected by some convenient relations between the real and imaginary parts of the functions $A_{1,2}$

$$A'_{1,2}(\omega_1)=A'_{1,2}(\omega_4),\ A''_{1,2}(\omega_1)=-A''_{1,2}(\omega_4),$$
$$A'_{1,2}(\omega_2)=A'_{1,2}(\omega_3),\ A''_{1,2}(\omega_2)=-A''_{1,2}(\omega_3).$$
(A16)

This permits to rewrite the sums in (A13) as follows

$$\mathrm{Re}\sum_{j=1}^{4}\frac{A_{1,2}(\omega_j)\mathrm{e}^{i\omega_j\tau}}{|D'(\omega_j)|}=\frac{2\mathrm{e}^{\gamma\tau}}{|D'(\omega_1)|}\left[A'_{1,2}(\omega_1)\cos\tilde{\omega}_1\tau-A''_{1,2}(\omega_4)\cos\tilde{\omega}_1\tau\right]$$
$$+\frac{2\mathrm{e}^{\gamma\tau}}{|D'(\omega_2)|}\left[A'_{1,2}(\omega_2)\cos\tilde{\omega}_2\tau-A''_{1,2}(\omega_3)\cos\tilde{\omega}_2\tau\right],$$
(A17)

Then, we can put additional restrictions $A'_2(\omega_1)=-A'_1(\omega_1)$, $A''_2(\omega_4)=-A''_1(\omega_4)$, and $A'_2(\omega_2)=A'_1(\omega_2)$, $A''_2(\omega_3)=A''_1(\omega_3)$, which follow from similar groundings. Other choice of relations leads to trivial identities.

Using the obtained relations we have the trajectories in the following forms

$$\xi_{p1}(\tau)=\mathrm{Re}\int_{-\infty}^{\infty}\frac{d\omega}{2\pi}C_1(\omega)\mathrm{e}^{i\omega\tau}+\frac{2\mathrm{e}^{\gamma\tau}}{|D'(\omega_1)|}\zeta_1(\tau)+\frac{2\mathrm{e}^{\gamma\tau}}{|D'(\omega_2)|}\zeta_2(\tau),$$
(A18)

$$\xi_{p2}(\tau) = \text{Re}\int_{-\infty}^{\infty}\frac{d\omega}{2\pi}C_2(\omega)e^{i\omega\tau} - \frac{2e^{\gamma\tau}}{|D'(\omega_1)|}\zeta_1(\tau) + \frac{2e^{\gamma\tau}}{|D'(\omega_2)|}\zeta_2(\tau),\quad\text{(A19)}$$

where

$$\begin{aligned}\zeta_1(\tau) &= A'_1(\omega_1)\cos\tilde{\omega}_1\tau - A''_1(\omega_4)\sin\tilde{\omega}_1\tau,\\ \zeta_2(\tau) &= A'_1(\omega_2)\cos\tilde{\omega}_2\tau - A''(\omega_3)\sin\tilde{\omega}_2\tau\end{aligned}\quad\text{(A21)}$$

Then, satisfying the conditions $\xi_{1,2}(0) = 0$ and $\xi_{1,2}(t) = 0$ we find the unknown functions $A'_1(\omega_1)$, $A'_1(\omega_2)$, $A''_1(\omega_3)$, $A''_1(\omega_4)$ and obtain the seeking for partial solutions

$$\begin{aligned}\xi_{p1}(\tau) &= \text{Re}\int_{-\infty}^{\infty}\frac{d\omega}{2\pi}C_1(\omega)e^{i\omega\tau} - \frac{e^{\gamma\tau-\gamma t}}{2}\left[\frac{\sin(\tilde{\omega}_1\tau)}{\sin(\tilde{\omega}_1 t)}C^-(t) + \frac{\sin(\tilde{\omega}_2\tau)}{\sin(\tilde{\omega}_2 t)}C^+(t)\right]\\ &\quad -\frac{e^{\gamma\tau}}{2}\left[\cos(\tilde{\omega}_1\tau) - \frac{\cos(\tilde{\omega}_1 t)}{\sin(\tilde{\omega}_1 t)}\sin(\tilde{\omega}_1\tau)\right]C^-(0) - \frac{e^{\gamma\tau}}{2}\left[\cos(\tilde{\omega}_2\tau) - \frac{\cos(\tilde{\omega}_2 t)}{\sin(\tilde{\omega}_2 t)}\sin(\tilde{\omega}_2\tau)\right]C^+(0)\end{aligned}\quad\text{(A22)}$$

$$\begin{aligned}\xi_{p2}(\tau) &= \text{Re}\int_{-\infty}^{\infty}\frac{d\omega}{2\pi}C_2(\omega)e^{i\omega\tau} + \frac{e^{\gamma\tau-\gamma t}}{2}\left[\frac{\sin(\tilde{\omega}_1\tau)}{\sin(\tilde{\omega}_1 t)}C^-(t) - \frac{\sin(\tilde{\omega}_2\tau)}{\sin(\tilde{\omega}_2 t)}C^+(t)\right]\\ &\quad +\frac{e^{\gamma\tau}}{2}\left[\cos(\tilde{\omega}_1\tau) - \frac{\cos(\tilde{\omega}_1 t)}{\sin(\tilde{\omega}_1 t)}\sin(\tilde{\omega}_1\tau)\right]C^-(0) - \frac{e^{\gamma\tau}}{2}\left[\cos(\tilde{\omega}_2\tau) - \frac{\cos(\tilde{\omega}_2 t)}{\sin(\tilde{\omega}_2 t)}\sin(\tilde{\omega}_2\tau)\right]C^+(0)\end{aligned}\quad\text{(A23)}$$

where

$$\begin{aligned}C^{\pm}(0) &= \text{Re}\int_{-\infty}^{\infty}\frac{d\omega}{2\pi}C_1(\omega) \pm \text{Re}\int_{-\infty}^{\infty}\frac{d\omega}{2\pi}C_2(\omega),\\ C^{\pm}(t) &= \text{Re}\int_{-\infty}^{\infty}\frac{d\omega}{2\pi}C_1(\omega)e^{i\omega t} \pm \text{Re}\int_{-\infty}^{\infty}\frac{d\omega}{2\pi}C_2(\omega)e^{i\omega t}.\end{aligned}\quad\text{(A24)}$$

Finally, we write a general solution of the homogeneous system of equations (A1) satisfying boundary conditions

$$\begin{aligned}\tilde{X}_1(\tau) &= w_1(t)\sin(\Omega_1\tau)\exp(-\delta_1\tau) + w_2\cos(\Omega_1\tau)\exp(-\delta_1\tau) +\\ &\quad + w_3(t)r_2\sin(\Omega_2\tau)\exp(-\delta_2\tau) + w_4 r_2\cos(\Omega_2\tau)\exp(-\delta_2\tau),\end{aligned}\quad\text{(A25)}$$

$$\begin{aligned}\tilde{X}_2(\tau) &= w_1(t)r_1\sin(\Omega_1\tau)\exp(-\delta_1\tau) + w_2 r_1\cos(\Omega_1\tau)\exp(-\delta_1\tau) +\\ &\quad + w_3(t)\sin(\Omega_2\tau)\exp(-\delta_2\tau) + w_4\cos(\Omega_2\tau)\exp(-\delta_2\tau),\end{aligned}\quad\text{(A26)}$$

where

$$w_1(t) = \left[\frac{X_{f1} - r_2 X_{f2}}{(1-r_1 r_2)\sin(\Omega_1 t)}\exp(\delta_1 t) - \cot(\Omega_1 t)\frac{X_{i1} - r_2 X_{i2}}{(1-r_1 r_2)}\right],\quad w_2 = \frac{X_{i1} - r_2 X_{i2}}{(1-r_1 r_2)},$$

$$w_3(t) = \left[\frac{X_{f2} - r_1 X_{f1}}{(1-r_1 r_2)\sin(\Omega_2 t)}\exp(\delta_2 t) - \cot(\Omega_2 t)\frac{X_{i2} - r_1 X_{i1}}{(1-r_1 r_2)}\right],\quad w_4 = \frac{X_{i2} - r_1 X_{i1}}{(1-r_1 r_2)}.\quad\text{(A27)}$$

A general solution of the inhomogeneous system of equation (A2) satisfying the conditions $\tilde{\xi}_{1,2}(0) = \xi_{i1,2}$ and $\tilde{\xi}_{1,2}(t) = \xi_{f1,2}$ is as follows

$$\tilde{\xi}_1(\tau) = v_1(t)\sin(\Omega_1\tau)\exp(\delta_1\tau) + v_2\cos(\Omega_1\tau)\exp(\delta_1\tau) + $$
$$+ v_3(t)r_2\sin(\Omega_2\tau)\exp(\delta_2\tau) + v_4 r_2\cos(\Omega_2\tau)\exp(\delta_2\tau) + \xi_{p1}(\tau), \quad (A28)$$

$$\tilde{\xi}_2(\tau) = v_1(t)r_1\sin(\Omega_1\tau)\exp(\delta_1\tau) + v_2 r_1\cos(\Omega_1\tau)\exp(\delta_1\tau) + $$
$$+ v_3(t)\sin(\Omega_2\tau)\exp(\delta_2\tau) + v_4\cos(\Omega_2\tau)\exp(\delta_2\tau) + \xi_{p2}(\tau), \quad (A29)$$

where

$$v_1(t) = \left[\frac{\xi_{f1} - r_2\xi_{f2}}{(1-r_1r_2)\sin(\Omega_1 t)}\exp(-\delta_1 t) - \cot(\Omega_1 t)\frac{\xi_{i1} - r_2\xi_{i2}}{(1-r_1r_2)}\right], \quad v_2 = \frac{\xi_{i1} - r_2\xi_{i2}}{(1-r_1r_2)}$$

$$v_3(t) = \left[\frac{\xi_{f2} - r_1\xi_{f1}}{(1-r_1r_2)\sin(\Omega_2 t)}\exp(-\delta_2 t) - \cot(\Omega_2 t)\frac{\xi_{i2} - r_1\xi_{i1}}{(1-r_1r_2)}\right], \quad v_4 = \frac{\xi_{i2} - r_1\xi_{i1}}{(1-r_1r_2)}, \quad (A30)$$

and partial solutions $\xi_{p1,p2}(\tau)$ are from (A22),(A23).

## Appendix B. Temporal functions related to Eq.(42)

Some of the temporal functions in Eq.(42) can be found in [45]. Here we repersent other functions which are necessary in calculating of a covariance matrix related to the case of driven coupled oscillators. Below we keep in mind that $g'(t) \equiv g'(t,\lambda,T_1,T_2)$, $g''(t) \equiv g''(t,\lambda,T_1,T_2)$, $a_{ext}^{\pm}(t) \equiv a_{ext}^{\pm}(t,\lambda,T_1,T_2)$, $b_{ext}^{\pm}(t) \equiv b_{ext}^{\pm}(t,\lambda,T_1,T_2)$, $A_{ext}^{\pm}(t) \equiv A_{ext}^{\pm}(t,\lambda,T_1,T_2)$, $B_{ext}^{\pm}(t) \equiv B_{ext}^{\pm}(t,\lambda,T_1,T_2)$ and, all the functions in Eq.(42) are pure real

$$g_1(t) = \frac{(D_9 + D_9' + \Pi_6)^2}{4\hbar(C_2 + \hbar a_2)} - \frac{e_6^2(C_2/\hbar + a_2)}{4\hbar a_2(C_2 + \hbar a_2) + (D_4' + \Pi_{16})^2} + \frac{Z_2^2}{4\hbar^2 Z_1} + \frac{Y_5^2}{4\hbar^2 Y_1},$$

$$g_{12}(t) = \frac{(D_3' + \Pi_8)(D_9 + D_9' + \Pi_6)}{2\hbar(C_2 + \hbar a_2)} - \frac{2e_5 e_6(C_2/\hbar + a_2)}{4\hbar a_2(C_2 + \hbar a_2) + (D_4' + \Pi_{16})^2} + \frac{Z_2 Z_3}{2\hbar^2 Z_1} + \frac{Y_4 Y_5}{2\hbar^2 Y_1}, \quad (B1)$$

$$g_2(t) = \frac{(D_3' + \Pi_8)^2}{4\hbar(C_2 + \hbar a_2)} - \frac{e_5^2(C_2/\hbar + a_2)}{4\hbar a_2(C_2 + \hbar a_2) + (D_4' + \Pi_{16})^2} + \frac{Z_3^2}{4\hbar^2 Z_1} + \frac{Y_4^2}{4\hbar^2 Y_1},$$

$$g_1'(t) = \frac{A_1}{\hbar} - \frac{E_3^2}{4\hbar(C_2 + \hbar a_2)} + \frac{e_2^2(C_2/\hbar + a_2)}{4\hbar a_2(C_2 + \hbar a_2) + (D_4' + \Pi_{16})^2} + \frac{Z_4^2}{4\hbar^2 Z_1} + \frac{Y_2^2}{4\hbar^2 Y_1},$$

$$g_{12}'(t) = \frac{E_4}{\hbar} - \frac{B_2 E_3}{2\hbar(C_2 + \hbar a_2)} + \frac{2e_1 e_2(C_2/\hbar + a_2)}{4\hbar a_2(C_2 + \hbar a_2) + (D_4' + \Pi_{16})^2} + \frac{Z_4 Z_5}{2\hbar^2 Z_1} + \frac{Y_2 Y_3}{2\hbar^2 Y_1}, \quad (B2)$$

$$g_2'(t) = \frac{A_2}{\hbar} - \frac{B_2^2}{4\hbar(C_2 + \hbar a_2)} + \frac{e_1^2(C_2/\hbar + a_2)}{4\hbar a_2(C_2 + \hbar a_2) + (D_4' + \Pi_{16})^2} + \frac{Z_5^2}{4\hbar^2 Z_1} + \frac{Y_3^2}{4\hbar^2 Y_1},$$

$$g''_{11}(t) = \frac{(D_1 + \Pi_1)}{\hbar} - \frac{E_3(D_9 + D'_9 + \Pi_6)}{2\hbar(C_2 + \hbar a_2)} - \frac{2e_2 e_6 (C_2/\hbar + a_2)}{4\hbar a_2(C_2 + \hbar a_2) + (D'_4 + \Pi_{16})^2} - \frac{Z_2 Z_4}{i2\hbar^2 Z_1} - \frac{Y_2 Y_5}{i2\hbar^2 Y_1},$$

$$g''_{21}(t) = \frac{(D_5 + D'_5 + \Pi_3)}{\hbar} - \frac{E_3(D'_3 + \Pi_8)}{2\hbar(C_2 + \hbar a_2)} - \frac{2e_2 e_5 (C_2/\hbar + a_2)}{4\hbar a_2(C_2 + \hbar a_2) + (D'_4 + \Pi_{16})^2} - \frac{Z_3 Z_4}{i2\hbar^2 Z_1} - \frac{Y_2 Y_4}{i2\hbar^2 Y_1},$$

$$g''_{12}(t) = \frac{(D_6 + D'_6 + \Pi_2)}{\hbar} - \frac{B_2(D_9 + D'_9 + \Pi_6)}{2\hbar(C_2 + \hbar a_2)} - \frac{2e_1 e_6 (C_2/\hbar + a_2)}{4\hbar a_2(C_2 + \hbar a_2) + (D'_4 + \Pi_{16})^2}$$
$$- \frac{Z_2 Z_5}{i2\hbar^2 Z_1} - \frac{Y_3 Y_5}{i2\hbar^2 Y_1}, \quad \text{(B3)}$$

$$g''_{22}(t) = \frac{(D'_1 + \Pi_4)}{\hbar} - \frac{B_2(D'_3 + \Pi_8)}{2\hbar(C_2 + \hbar a_2)} - \frac{2e_1 e_5 (C_2/\hbar + a_2)}{4\hbar a_2(C_2 + \hbar a_2) + (D'_4 + \Pi_{16})^2} - \frac{Z_3 Z_5}{i2\hbar^2 Z_1} - \frac{Y_3 Y_4}{i2\hbar^2 Y_1}.$$

The function which take into account of external forces are

$$a^+_{ext}(t) = \frac{(D_9 + D'_9 + \Pi_6)\Lambda_2}{2\hbar(C_2 + \hbar a_2)} + \frac{(-e_6 \Lambda_2 (D'_4 + \Pi_{16}))}{\hbar[4\hbar a_2(C_2 + \hbar a_2) + (D'_4 + \Pi_{16})^2]}$$
$$+ \frac{Z_2(\Lambda_1 - \Lambda_2[On])}{2\hbar^2 Z_1} + \frac{(-y_5 \Lambda_2[Ro])}{2\hbar^2 Y_1} + \frac{(-y_5 \Lambda_1 Z_6)}{4\hbar^3 Y_1 Z_1},$$

$$b^+_{ext}(t) = \frac{(D'_3 + \Pi_8)\Lambda_2}{2\hbar(C_2 + \hbar a_2)} + \frac{(-e_5 \Lambda_2 (D'_4 + \Pi_{16}))}{\hbar[4\hbar a_2(C_2 + \hbar a_2) + (D'_4 + \Pi_{16})^2]} \quad \text{(B4)}$$
$$+ \frac{Z_3(\Lambda_1 - \Lambda_2[On])}{2\hbar^2 Z_1} + \frac{(-y_4 \Lambda_2[Ro])}{2\hbar^2 Y_1} + \frac{(-y_4 \Lambda_1 Z_6)}{4\hbar^3 Y_1 Z_1},$$

$$A^-_{ext}(t) = \frac{1}{\hbar}\left[\frac{M_1(t) + r_1 M'_1(t)}{(1 - r_1 r_2)\sin(\Omega_1 t)\exp(\delta_1 t)} - \frac{r_1 r_2 M_2(t) + r_1 M'_2(t)}{(1 - r_1 r_2)\sin(\Omega_2 t)\exp(\delta_2 t)}\right] + \frac{(-E_3 \Lambda_2)}{2\hbar(C_2 + \hbar a_2)}$$
$$+ \frac{z_4(\Lambda_2[On] - \Lambda_1)}{2\hbar^2 Z_1} + \frac{(-e_2 \Lambda_2)(D'_4 + \Pi_{16})}{\hbar[4\hbar a_2(C_2 + \hbar a_2) + (D'_4 + \Pi_{16})^2]} + \frac{(-Y_2)}{2\hbar^2 Y_1}(\Lambda_2[Ro] + \Lambda_1 \frac{Z_6}{2\hbar Z_1}),$$

$$B^-_{ext}(t) = \frac{1}{\hbar}\left[\frac{r_2 M_2(t) + M'_2(t)}{(1 - r_1 r_2)\sin(\Omega_2 t)\exp(\delta_2 t)} - \frac{r_1 r_2 M'_1(t) + r_2 M_1(t)}{(1 - r_1 r_2)\sin(\Omega_1 t)\exp(\delta_1 t)}\right] + \frac{(-B_2 \Lambda_2)}{2\hbar(C_2 + \hbar a_2)} \quad \text{(B5)}$$
$$+ \frac{z_5(\Lambda_2[On] - \Lambda_1)}{2\hbar^2 Z_1} + \frac{(-e_1 \Lambda_2)(D'_4 + \Pi_{16})}{\hbar[4\hbar a_2(C_2 + \hbar a_2) + (D'_4 + \Pi_{16})^2]} + \frac{(-Y_3)}{2\hbar^2 Y_1}(\Lambda_2[Ro] + \Lambda_1 \frac{Z_6}{2\hbar Z_1}),$$

where $On = \dfrac{E_1}{2(C_2 + \hbar a_2)} + \dfrac{e_3(D'_4 + \Pi_{16})}{4\hbar a_2(C_2 + \hbar a_2) + (D'_4 + \Pi_{16})^2}$, and

$$Ro = \frac{(D_{11} + D'_{11} + \Pi_{14})}{2(C_2 + \hbar a_2)} - \frac{e_4(D'_4 + \Pi_{16})}{4\hbar a_2(C_2 + \hbar a_2) + (D'_4 + \Pi_{16})^2} - [On]\frac{Z_6}{2\hbar Z_1}.$$

$$\Phi(t) = \frac{(\xi_{f1} - r_2 \xi_{f2})[M_1(t) + r_1 M'_1(t)]}{(1 - r_1 r_2)\sin(\Omega_1 t)\exp(\delta_1 t)} + \frac{(\xi_{f2} - r_1 \xi_{f1})[M'_2(t) + r_2 M_2(t)]}{(1 - r_1 r_2)\sin(\Omega_2 t)\exp(\delta_2 t)}, \quad \text{(B6)}$$

$$\Lambda_1(t) = \frac{1}{(1-r_1r_2)} \{[-cot(\Omega_1 t)M_1(t) + N_1(t) + r_1r_2 cot(\Omega_2 t)M_2(t) - r_1r_2 N_2(t)] +$$

$$+ [-r_1 cot(\Omega_1 t)M_1'(t) + r_1 N_1'(t) + r_1 cot(\Omega_2 t)M_2'(t) - r_1 N_2'(t)]\},$$

$$\Lambda_2(t) = \frac{1}{(1-r_1r_2)} \{[r_2 cot(\Omega_1 t)M_1(t) - r_2 N_1(t) - r_2 cot(\Omega_2 t)M_2(t) + r_2 N_2(t)] + \quad \text{(B7)}$$

$$+ [r_1r_2 cot(\Omega_1 t)M_1'(t) - r_1r_2 N_1'(t) - cot(\Omega_2 t)M_2'(t) + N_2'(t)]\},$$

where

$$M_k(t) = \int_0^t d\tau \, f_1(\tau) \sin(\Omega_k \tau) exp(\delta_k \tau),$$

$$N_k(t) = \int_0^t d\tau \, f_1(\tau) \cos(\Omega_k \tau) exp(\delta_k \tau), \ (k=1,2)$$

$$M_k'(t) = \int_0^t d\tau \, f_2(\tau) \sin(\Omega_k \tau) exp(\delta_k \tau), \quad \text{(B8)}$$

$$N_k'(t) = \int_0^t d\tau \, f_2(\tau) \cos(\Omega_k \tau) exp(\delta_k \tau), \ (k=1,2)$$

Below the functions which are not involved in our numerical calculations, but they determine the non-Hermitian matrix in Eq. (40).

$$a_{ext}^-(t) = \frac{2e_2 V_2(C_2 + \hbar a_2)}{\hbar[4\hbar a_2(C_2 + \hbar a_2) + (D_4' + \Pi_{16})^2]} + \frac{(-z_4 V_2)}{\hbar^2 Z_1} \frac{e_3(D_4' + \Pi_{16})}{4\hbar a_2(C_2 + \hbar a_2) + (D_4' + \Pi_{16})^2}$$

$$+ \frac{Y_2}{2\hbar^2 Y_1} [V_1 - 2V_2 \frac{(e_4 + e_3 Z_6 / 2\hbar Z_1)(C_2 + \hbar a_2)}{4\hbar a_2(C_2 + \hbar a_2) + (D_4' + \Pi_{16})^2}],$$

$$b_{ext}^-(t) = \frac{2e_1 V_2(C_2 + \hbar a_2)}{\hbar[4\hbar a_2(C_2 + \hbar a_2) + (D_4' + \Pi_{16})^2]} + \frac{(-z_5 V_2)}{\hbar^2 Z_1} \frac{e_3(D_4' + \Pi_{16})}{4\hbar a_2(C_2 + \hbar a_2) + (D_4' + \Pi_{16})^2} \quad \text{(B9)}$$

$$+ \frac{Y_3}{2\hbar^2 Y_1} [V_1 - 2V_2 \frac{(e_4 + e_3 Z_6 / 2\hbar Z_1)(C_2 + \hbar a_2)}{4\hbar a_2(C_2 + \hbar a_2) + (D_4' + \Pi_{16})^2}],$$

$$A_{ext}^+(t) = \frac{(-2e_6 V_2(C_2 + \hbar a_2))}{\hbar[4\hbar a_2(C_2 + \hbar a_2) + (D_4' + \Pi_{16})^2]} + \frac{(-Z_2 V_2)}{\hbar^2 Z_1} \frac{e_3(D_4' + \Pi_{16})}{4\hbar a_2(C_2 + \hbar a_2) + (D_4' + \Pi_{16})^2}$$

$$+ \frac{(-y_5)}{2\hbar^2 Y_1} [V_1 - 2V_2 \frac{(e_4 + e_3 Z_6 / 2\hbar Z_1)(C_2 + \hbar a_2)}{4\hbar a_2(C_2 + \hbar a_2) + (D_4' + \Pi_{16})^2}];$$

$$B_{ext}^+(t) = \frac{(-2e_5 V_2(C_2 + \hbar a_2))}{\hbar[4\hbar a_2(C_2 + \hbar a_2) + (D_4' + \Pi_{16})^2]} + \frac{(-Z_3 V_2)}{\hbar^2 Z_1} \frac{e_3(D_4' + \Pi_{16})}{4\hbar a_2(C_2 + \hbar a_2) + (D_4' + \Pi_{16})^2} \quad \text{(B10)}$$

$$+ \frac{(-y_4)}{2\hbar^2 Y_1} [V_1 - 2V_2 \frac{(e_4 + e_3 Z_6 / 2\hbar Z_1)(C_2 + \hbar a_2)}{4\hbar a_2(C_2 + \hbar a_2) + (D_4' + \Pi_{16})^2}];$$

Juxtaposition of $a_{ext}^-(t)$, $b_{ext}^-(t)$, $A_{ext}^+(t)$, $B_{ext}^+(t)$ with $a_{ext}^+(t)$, $b_{ext}^+(t)$, $A_{ext}^-(t)$, $B_{ext}^-(t)$ shows that in most cases we must eventually compare the functions $V_{1,2}$ with $\Lambda_{1,2}$. Our verification shows that $V_{1,2} \sim 10^{-21} \Lambda_{1,2}$, and we did not write out them here, because their irrelevance.

# Appendix C. Mean values, variances and covariances of coordinates and momenta of driven coupled oscillators

For further convenience we introduce designations $\beta_{11} = 8g_1(t)$, $\beta_{12} = 4g_{12}(t)$, $\beta_{22} = 8g_2(t)$, and $A^-_{ext}(t) = A$, $B^-_{ext}(t) = B$, $a^+_{ext}(t) = a$, $b^+_{ext}(t) = b$ for brevity.

Then the total Hermitian density matrix with external forces is

$$
\begin{aligned}
\rho(X_{f1}, X_{f2}, t; \xi_{f1}, \xi_{f2}, 0) &= C(t) \exp\left\{-\left[\beta_{11}(t) X_{f1}^2/8 + \beta_{12}(t) X_{f1} X_{f2}/4 + \beta_{22}(t) X_{f2}^2/8\right]\right\} \\
&\times \exp\left\{-\left[g_1'(t)\xi_{f1}^2 + g_{12}'(t)\xi_{f1}\xi_{f2} + g_2'(t)\xi_{f2}^2\right]\right\} \\
&\times \exp\left\{i\left[g_{11}''(t) X_{f1}\xi_{f1} + g_{21}''(t) X_{f2}\xi_{f1} + g_{12}''(t) X_{f1}\xi_{f2} + g_{22}''(t) X_{f2}\xi_{f2}\right]\right\} \\
&\times \exp\left\{iA^-_{ext}(t)\xi_{f1} + iB^-_{ext}(t)\xi_{f2}\right\} \exp\left\{-a^+_{ext}(t) X_{f1} - b^+_{ext}(t) X_{f2}\right\}
\end{aligned}
\tag{C1}
$$

where $X_{f\alpha} = x_{f\alpha} + y_{f\alpha}$ and $\xi_{f\alpha} = x_{f\alpha} - y_{f\alpha}$, ($\alpha = 1, 2$).

It is known that in order to obtain mean values of observables of one oscillator we should to reduce the total density matrix with respect to the observables of other oscillator. For example

$$
\rho(x_{f1}, t; y_{f1}, 0) = \int_{-\infty}^{\infty} \lim_{y_{f2} \to x_{f2}} \{\rho(x_{f1}, x_{f2}, t; y_{f1}, y_{f2}, 0)\} dx_{f2} , \tag{C2}
$$

or with use of the other variables

$$
\rho(X_{f1}, t; \xi_{f1}, 0) = \int_{-\infty}^{\infty} \lim_{\xi_{f2} \to 0} \{\rho(X_{f1}, X_{f2}, t; \xi_{f1}, \xi_{f2}, 0)\} dX_{f2} , \tag{C3}
$$

and a similar formulae for the second oscillator.

The reduced density matrixes are as follows

$$
\begin{aligned}
\rho(X_{f1}, t; \xi_{f1}, 0) = C\sqrt{\frac{8\pi}{\beta_{22}}} &\exp[-aX_{f1} - (\beta_{11}/8) X_{f1}^2 + iA\xi_{f1} + ig_{11}'' X_{f1}\xi_{f1} \\
&+ (4b + \beta_{12} X_{f1} - 4ig_{21}''\xi_{f1})^2/8\beta_{22} - g_1'\xi_{f1}^2],
\end{aligned}
\tag{C4}
$$

$$
\begin{aligned}
\rho(X_{f2}, t; \xi_{f2}, 0) = C\sqrt{\frac{8\pi}{\beta_{11}}} &\exp[-aX_{f2} - (\beta_{22}/8) X_{f2}^2 + iB\xi_{f2} + ig_{22}'' X_{f2}\xi_{f2} \\
&+ (4a + \beta_{12} X_{f2} - 4ig_{12}''\xi_{f2})^2/8\beta_{11} - g_2'\xi_{f2}^2].
\end{aligned}
\tag{C5}
$$

The normalisation constant $C(t)$ can be found, for instance from Eq.(C4) or Eq.(C5)

$$
1 = \int_{-\infty}^{\infty} \lim_{\xi_{f1} \to 0} \{\rho(X_{f1}, t; \xi_{f1}, 0)\} dX_{f1} = \int_{-\infty}^{\infty} \lim_{\xi_{f2} \to 0} \{\rho(X_{f2}, t; \xi_{f2}, 0)\} dX_{f2} . \tag{C6}
$$

For cross-correlated values we need another normalisation condition as follows

$$
1 = \int_{-\infty}^{\infty} \int_{-\infty}^{\infty} \lim_{\substack{\xi_{f1} \to 0 \\ \xi_{f2} \to 0}} \{\rho(X_{f1}, X_{f2}, t; \xi_{f1}, \xi_{f2}, 0)\} dX_{f1} dX_{f2} , \tag{C7}
$$

For any case the normalisation constant is

$$C(t) = \frac{\sqrt{\beta_{11}\beta_{22} - \beta_{12}^2}}{4\pi} \exp\left[-\frac{2(b^2\beta_{11} - 2ab\beta_{12} + a^2\beta_{22})}{\beta_{11}\beta_{22} - \beta_{12}^2}\right]. \tag{C8}$$

After this we have everything to calculate all the mean values, variances and covariances as usual

$$\begin{aligned}
\bar{L}_1 &= \int_{-\infty}^{\infty} \lim_{y_{f1} \to x_{f1}} \{\hat{L}_1 \rho(x_{f1}, t; y_{f1}, 0)\} dx_{f1}, \\
\bar{L}_2 &= \int_{-\infty}^{\infty} \lim_{y_{f2} \to x_{f2}} \{\hat{L}_2 \rho(x_{f2}, t; y_{f2}, 0)\} dx_{f2}, \\
\bar{L}_{mn} &= \int_{-\infty}^{\infty}\int_{-\infty}^{\infty} \lim_{\substack{y_{f1} \to x_{f1} \\ y_{f2} \to x_{f2}}} \{\hat{L}_{mn} \rho(x_{f1}, x_{f2}, t; y_{f1}, y_{f2}, 0)\} dx_{f1} dx_{f2}, \quad (m, n = 1, 2)
\end{aligned} \tag{C9}$$

where $\hat{L}_\alpha = \{\hat{x}_\alpha, \hat{p}_\alpha, (\hat{x}_\alpha)^2, (\hat{p}_\alpha)^2\}, (\alpha = 1, 2)$, and $\hat{L}_{mn} = \{\hat{x}_m\hat{x}_n, \hat{p}_m\hat{p}_n, \hat{x}_m\hat{p}_n, \hat{p}_m\hat{x}_n\}, (m, n = 1, 2)$, where $\hat{p}_m = -i\hbar\partial/\partial x_m$.

Using above written formulas we find all the characteristics.

For the first oscillator the mean value of the coordinate

$$\bar{x}_1(t) = \frac{2(b\beta_{12} - a\beta_{22})}{\beta_{11}\beta_{22} - \beta_{12}^2}, \tag{C10}$$

for the mean value of the coordinate in square

$$\overline{x_1^2}(t) = \frac{4b^2\beta_{12}^2 - \beta_{12}\beta_{22}(8ab + \beta_{12}) + (4a^2 + \beta_{11})\beta_{22}^2}{(\beta_{11}\beta_{22} - \beta_{12}^2)^2}, \tag{C11}$$

for the dispersion of the coordinate

$$\overline{x_1^2}(t) - \bar{x}_1^2(t) = \frac{\beta_{22}}{\beta_{11}\beta_{22} - \beta_{12}^2}, \tag{C12}$$

for the mean value of the momentum

$$\bar{p}_1(t) = 4\hbar \frac{a(g''_{21}\beta_{12} - g''_{11}\beta_{22}) + b(g''_{11}\beta_{12} - g''_{21}\beta_{11})}{\beta_{11}\beta_{22} - \beta_{12}^2} + \hbar A, \tag{C13}$$

for the mean value of the momentum in square

$$\begin{aligned}
\overline{p_1^2}(t) = \hbar^2(\beta_{11}\beta_{22} - \beta_{12}^2)^{-2}\{&16b^2(g''_{21}\beta_{11} - g''_{11}\beta_{12})^2 + 16a^2(g''_{21}\beta_{12} - g''_{11}\beta_{22})^2 \\
&- 8aA(g''_{21}\beta_{12} - g''_{11}\beta_{22})(\beta_{12}^2 - \beta_{11}\beta_{22}) - 8b(g''_{21}\beta_{11} - g''_{11}\beta_{12}) \times \\
&\times [4a(g''_{21}\beta_{12} - g''_{11}\beta_{22}) + A(\beta_{11}\beta_{22} - \beta_{12}^2)] + (\beta_{12}^2 - \beta_{11}\beta_{22})[8g''_{11}g''_{21}\beta_{12} \\
&- 4g''^2_{21}\beta_{11} - 4g''^2_{11}\beta_{22} + A^2(\beta_{12}^2 - \beta_{11}\beta_{22})] + 2(\beta_{12}^2 - \beta_{11}\beta_{22})^2 g'_1\}
\end{aligned} \tag{C14}$$

for the dispersion of the momentum

$$\overline{p_1^2}(t) - \bar{p}_1^2(t) = 2\hbar^2\left[\frac{2(g''^2_{21}\beta_{11} - 2g''_{11}g''_{21}\beta_{12} + g''^2_{11}\beta_{22})}{\beta_{11}\beta_{22} - \beta_{12}^2} + g'_1\right], \tag{C15}$$

for the simmetrised mean value of the cross covariance for the first oscillator

$$[\overline{x_1 p_1}(t) + \overline{p_1 x_1}(t)]/2 = 2\hbar(\beta_{11}\beta_{22} - \beta_{12}^2)^{-2}\{4b^2\beta_{12}(g''_{11}\beta_{12} - g''_{21}\beta_{11})$$
$$+4a^2\beta_{22}(g''_{11}\beta_{22} - g''_{21}\beta_{12}) + (g''_{11}\beta_{22} - g''_{21}\beta_{12})(\beta_{11}\beta_{22} - \beta_{12}^2) + bA(\beta_{11}\beta_{22}\beta_{12} - \beta_{12}^3) \,, \quad \text{(C16)}$$
$$+b[A\beta_{12}(\beta_{12}^2 - \beta_{11}\beta_{22}) + 4a(g''_{21}\beta_{12}^2 - 2g''_{11}\beta_{12}\beta_{22} + g''_{21}\beta_{11}\beta_{22})]\}$$

for the commutator of the first oscillator

$$\overline{x_1 p_1}(t) - \overline{p_1 x_1}(t) = i\hbar \,. \qquad \text{(C17)}$$

For the second oscillator the mean value of the coordinate

$$\overline{x_2}(t) = \frac{2(a\beta_{12} - b\beta_{11})}{\beta_{11}\beta_{22} - \beta_{12}^2} \,, \qquad \text{(C18)}$$

for the mean value of the coordinate in square

$$\overline{x_2^2}(t) = \frac{4a^2\beta_{12}^2 - \beta_{12}\beta_{11}(8ab + \beta_{12}) + (4b^2 + \beta_{22})\beta_{11}^2}{(\beta_{11}\beta_{22} - \beta_{12}^2)^2} \,, \qquad \text{(C19)}$$

for the dispersion of the coordinate

$$\overline{x_2^2}(t) - \overline{x_2}^2(t) = \frac{\beta_{11}}{\beta_{11}\beta_{22} - \beta_{12}^2} \,, \qquad \text{(C20)}$$

for the mean value of the momentum

$$\overline{p_2}(t) = 4\hbar \frac{a(g''_{12}\beta_{22} - g''_{22}\beta_{12}) + b(g''_{22}\beta_{11} - g''_{12}\beta_{12})}{\beta_{11}\beta_{22} - \beta_{12}^2} + \hbar B \,, \qquad \text{(C21)}$$

for the mean value of the momentum in square

$$\overline{p_2^2}(t) = \hbar^2(\beta_{11}\beta_{22} - \beta_{12}^2)^{-2}\{16b^2(g''_{22}\beta_{11} - g''_{12}\beta_{12})^2 + 16a^2(g''_{22}\beta_{12} - g''_{12}\beta_{22})^2$$
$$-8aB(g''_{22}\beta_{12} - g''_{12}\beta_{22})(\beta_{12}^2 - \beta_{11}\beta_{22}) - 8b(g''_{22}\beta_{11} - g''_{12}\beta_{12}) \times$$
$$\times[4a(g''_{22}\beta_{12} - g''_{12}\beta_{22}) + B(\beta_{11}\beta_{22} - \beta_{12}^2)] + (\beta_{12}^2 - \beta_{11}\beta_{22})[8g''_{12}g''_{22}\beta_{12} \qquad , \quad \text{(C22)}$$
$$-4g''^2_{22}\beta_{11} - 4g''^2_{12}\beta_{22} + B^2(\beta_{12}^2 - \beta_{11}\beta_{22})] + 2(\beta_{12}^2 - \beta_{11}\beta_{22})^2 g'_2\}$$

for the dispersion of the momentum

$$\overline{p_2^2}(t) - \overline{p_2}^2(t) = 2\hbar^2\left[\frac{2(g''^2_{22}\beta_{11} - 2g''_{12}g''_{22}\beta_{12} + g''^2_{12}\beta_{22})}{\beta_{11}\beta_{22} - \beta_{12}^2} + g'_2\right] \,, \qquad \text{(C23)}$$

for the simmetrised mean value of the cross covariance for the second oscillator

$$[\overline{x_2 p_2}(t) + \overline{p_2 x_2}(t)]/2 = 2\hbar(\beta_{11}\beta_{22} - \beta_{12}^2)^{-2}\{4b^2\beta_{11}(g''_{22}\beta_{11} - g''_{12}\beta_{12})$$
$$+4a^2\beta_{12}(g''_{22}\beta_{12} - g''_{12}\beta_{22}) + (g''_{22}\beta_{11} - g''_{12}\beta_{12})(\beta_{11}\beta_{22} - \beta_{12}^2) + aB(\beta_{11}\beta_{22}\beta_{12} - \beta_{12}^3) \,, \quad \text{(C24)}$$
$$+b[B\beta_{11}(\beta_{12}^2 - \beta_{11}\beta_{22}) + 4a(g''_{12}\beta_{12}^2 - 2g''_{22}\beta_{11}\beta_{12} + g''_{12}\beta_{11}\beta_{22})]\}$$

for the commutator of the second oscillator

$$\overline{x_2 p_2}(t) - \overline{p_2 x_2}(t) = i\hbar \,, \qquad \text{(C25)}$$

for the coordinate covariance of both oscillators

$$\overline{x_1 x_2}(t) = \frac{4ab(\beta_{12}^2 + \beta_{11}\beta_{22}) + \beta_{12}^3 - \beta_{12}\beta_{22}(4a^2 + \beta_{11}) - 4b^2\beta_{11}\beta_{12}}{(\beta_{11}\beta_{22} - \beta_{12}^2)^2} ,\quad (C26)$$

for the momentum covariance of both oscillators

$$\overline{p_1 p_2}(t) = \overline{p_2 p_1}(t) = 0 ,\quad (C27)$$

for the other covariances of both oscillators

$$\overline{x_1 p_2}(t) = \overline{p_2 x_1}(t) = \overline{x_2 p_1}(t) = \overline{p_1 x_2}(t) = 0 .\quad (C28)$$

It is easy to obtain from above written formulas all elements of the covariance matrix for the case of zero external forces by putting $A = B = a = b = 0$. In this case all the mean values are equal zero $\overline{x}_1(t) = \overline{x}_2(t) = \overline{p}_1(t) = \overline{p}_2(t) = 0$. That is why the formulas have a simpler view.

Namely, for the dispersion of the coordinate of the first oscillator

$$\overline{x_1^2}(t) = \frac{\beta_{22}}{\beta_{11}\beta_{22} - \beta_{12}^2} ,\quad (C29)$$

for the dispersion of the momentum of the first oscillator

$$\overline{p_1^2}(t) = 2\hbar^2 \left[ \frac{2(g_{21}''^{\,2}\beta_{11} - 2g_{11}''g_{21}''\beta_{12} + g_{11}''^{\,2}\beta_{22})}{\beta_{11}\beta_{22} - \beta_{12}^2} + g_1' \right] ,\quad (C30)$$

for the simmetrised mean value of the cross covariance for the first oscillator

$$[\overline{x_1 p_1}(t) + \overline{p_1 x_1}(t)]/2 = 2\hbar \frac{(g_{22}''\beta_{11} - g_{21}''\beta_{12})}{\beta_{11}\beta_{22} - \beta_{12}^2} ,\quad (C31)$$

for the commutator of the first oscillator

$$\overline{x_1 p_1}(t) - \overline{p_1 x_1}(t) = i\hbar ,\quad (C32)$$

for the dispersion of the coordinate of the second oscillator

$$\overline{x_2^2}(t) = \frac{\beta_{11}}{\beta_{11}\beta_{22} - \beta_{12}^2} ,\quad (C33)$$

for the dispersion of the momentum of the second oscillator

$$\overline{p_2^2}(t) = 2\hbar^2 \left[ \frac{2(g_{22}''^{\,2}\beta_{11} - 2g_{12}''g_{22}''\beta_{12} + g_{12}''^{\,2}\beta_{22})}{\beta_{11}\beta_{22} - \beta_{12}^2} + g_2' \right] ,\quad (C34)$$

for the simmetrised mean value of the cross covariance for the second oscillator

$$[\overline{x_2 p_2}(t) + \overline{p_2 x_2}(t)]/2 = 2\hbar \frac{(g_{22}''\beta_{11} - g_{12}''\beta_{12})}{\beta_{11}\beta_{22} - \beta_{12}^2} ,\quad (C33)$$

for the coordinate covariance of both oscillators

$$\overline{x_1 x_2}(t) = \frac{\beta_{12}}{\beta_{11}\beta_{22} - \beta_{12}^2} ,\quad (C35)$$

for the commutator of the second oscillator

$$\overline{x_2 p_2}(t) - \overline{p_2 x_2}(t) = i\hbar , \qquad (C36)$$

and other covariances of both oscillators are equal zero.

Finally, we have total set of 16 matrix elements for the $4\times 4$ covariance matrix both in cases with and without external forces and mean values of observables.

**Figure captions:**

**Figure 1.** A sketch of the problem studied: a) - two independent reservoirs 1, 2, and two uncoupled selected oscillators; b) - the couplings among all oscillators are switched on at $t \geq 0$; c) –the first external force is applied to the first coupled oscillator at $t = t_{01}$; d) –the second external force is applied to the second coupled oscillator at $t = t_{02}$.

**Figure 2.** Normalized mean values of the coordinates $\bar{x}_i(t)/\sqrt{\sigma_{0i}^2}$, $(i=1,2)$ of the first –a) and second –b) oscillators versus a time $t$, where $\sigma_{0i}^2 = \hbar/2M_i\omega_{0i}$, $(i=1,2)$ using Eq.(C10, C18). This figures correspond to the case of no coupling $\lambda = 0$, and total equilibrium, when $T_1 = T_2 = 300K$. The temporal dynamics of external forces $f_{1,2}(t)$ in accordance with Eq.(46) is shown between of these graphs.

**Figure 3.** Normalized mean values of the coordinates $\bar{x}_i(t)/\sqrt{\sigma_{0i}^2}$, $(i=1,2)$ of the first –a) and second –b) oscillators versus a time $t$ using Eq.(C10, C18) at relatively strong coupling $\tilde{\lambda} = \lambda/\omega_{01}\omega_{02}\sqrt{M_1 M_2} = 0.3$ between of oscillators. These figures also correspond to the case of the total equilibrium, when $T_1 = T_2 = 300K$.

**Figure 4.** Normalized mean values of the coordinates $\bar{x}_i(t)/\sqrt{\sigma_{0i}^2}$, $(i=1,2)$ of the first –a) and second –b) oscillators versus a time $t$ using Eq.(C10, C18) at $\tilde{\lambda} = 0.3$ between of oscillators in the case out of total equilibrium in the system, when $T_1 = 300K$ and $T_2 = 900K$.